\documentclass[
	aps
	,a4paper
	,pra
	,eqsecnum
	,showpacs
	,floatfix
	,twocolumn
	,merge
	,nofootinbib
	,elide
			]{revtex4-2}
\usepackage{amsmath}
\usepackage{amsfonts}		
\usepackage{graphicx}		
\usepackage{tikz}			
\usepackage{fourier}		
\usepackage{amssymb}

\usepackage[usenames,dvipsnames,svgnames]{pstricks}
\usepackage{soul}
\setul{}{1.5pt}
\setstcolor{red}

\definecolor{morange}{rgb}{1.0, 0.31, 0.0}

\usepackage[%
	unicode=true
	,colorlinks=true
	,linkcolor=blue
	,anchorcolor=blue
	,breaklinks=true
	,citecolor=blue
	,filecolor=cyan
	,menucolor=blue
	,breaklinks=true
	,urlcolor=blue
			]{hyperref}

\begin{document}
\title{Molecular magnetism in the multi-configurational self-consistent field method}

\author{M.~Georgiev}
\email{mgeorgiev@issp.bas.bg}
\affiliation{Institute of Solid State Physics, Bulgarian Academy of Sciences,
	Tsarigradsko Chauss\'ee 72, 1784 Sofia, Bulgaria}

\author{H.~Chamati}
\email{chamati@issp.bas.bg}
\affiliation{Institute of Solid State Physics, Bulgarian Academy of Sciences,
	Tsarigradsko Chauss\'ee 72, 1784 Sofia, Bulgaria}
	
\date{\today}
\begin{abstract}
We develop a structured theoretical framework used in
our recent articles [\href{https://doi.org/10.1140/epjb/e2019-100115-1}{Eur. Phys. J. B \textbf{92}, 93 (2019)}
and \href{https://doi.org/10.1103/PhysRevB.101.094427}{Phys. Rev. B
\textbf{101}, 094427 (2020)}]
to characterize the unusual behavior of the magnetic
spectrum, magnetization and magnetic susceptibility
of the molecular magnet Ni$_4$Mo$_{12}$.
The theoretical background is based on the molecular orbital theory in
conjunction with the multi-configurational self-consistent field method and 
results in a post-Hartree-Fock scheme for constructing the
corresponding  energy spectrum.
Furthermore, we
construct a bilinear spin-like Hamiltonian involving discrete
coupling parameters accounting for the relevant spectroscopic
magnetic excitations, magnetization and magnetic
susceptibility. The explicit expressions of the eigenenergies
of the ensuing Hamiltonian are determined and the physical
origin of broadening and splitting of experimentally observed
peaks in the magnetic spectra is discussed. 
To demonstrate the efficiency of our method we compute 
the spectral properties of a spin-one magnetic dimer.
The present approach may be applied to a variety of magnetic
units based on transition metals and rare earth elements. 
\end{abstract}
\maketitle
	

\section{Introduction}\label{sec:intro}

Since decades molecular magnets have secured their
own place as prominent tools for gaining insights into
the origin of magnetic exchange phenomena \cite{gatteschi_molecular_2006,winpenny_molecular_2011,bartolome_molecular_2014,gao_molecular_2015,coronado_molecular_2020} 
and studying magneto-related effects \cite{liu_electric_2019,vyaselev_spin_2020,cornia_2020,kowalewska_magnetocaloric_2020}.
The magnetic properties of such small in size and isolated
molecular units of matter are inevitably related to their size, symmetry, number 
and types of chemical bonds (for more details see
\cite{chamati_interaction_2010,chamati_theory_2013,sellmyer_novel_2015,sieklucka_molecular_2017}
and references therein). On the microscopic level, the low-spin 
and short-bridged magnetic compounds are ideal quantum systems for studying 
the origin of magneto-structural dependencies.
Dimeric \cite{hay_1975,felthouse_1977,hart_estimation_1992,guo_strong_2017}
and trimeric
\cite{gehring_1993,yoon_electronic_2007,ferrer_antisymmetric_2012}
magnetic units are among
the first and most widely explored sorts of spin clusters.
In that regard, cooper based compounds have served as
primordial tools in revealing the 
interdependence between exchange and structure parameters 
\cite{astheimer_1986,charlot_1986,lorosch_1987,gehring_1993,aebersold_1998}.
It is worth mentioning that, the relationship between
bridging composition and magnetic exchange phenomena may be attributed to the 
study of different ligand bridged complexes with a variety of magnetic centers 
\cite{ribas_1993,ribas_1999,song_unusual_2004,sadhu_2017,zhao_2017,fraser_2017}.
Other prominent investigations related to the magneto-structural effects are complexes
with Fe magnetic centers 
\cite{angaridis_2005,mekuimemba_2018,gregoli_2009,viennois_2010,kuzmann_magnetic_2017}, 
 Ni based 
compounds \cite{schnack_observation_2006,nehrkorn_inelastic_2010,loose_2008, panja_2017,das_2017,woods_2017} and 
Mn spin clusters 
\cite{goodenough_1955,defotis_1990,law_2000,han_2004,perks_2012,gupta_2016,hanninen_2018}.

For many isolated magnetic units that exhibit a mono or
diatomic intermediate structure the ensuing
exchange interactions may be approximately described by considering
density functional theory, Hartree-Fock or non-sophisticated
post-Hartree-Fock methods \cite{mcquarrie_quantum_2008,magnasco_2013,post-hartree-fock_2019}. The electrons
may be considered as localized. Thus the
Goodenough-Kanamori-Anderson rules hold and the hopping and single-ion coulomb repulsion terms
represent the leading electrons' correlations.
For such trivial bridging structure the analytical analysis allows the application of
conventional microscopic spin models. 
Of particular interest are the Heisenberg and Hubbard models, since
they possess just a few
parameters to capture the main isotropic magnetic futures.
On the other hand, any orbital contribution of the localized electrons into the exchange processes
may require involving
an anisotropic spin bilinear expression and/or including a single-ion anisotropy term
accounting for the zero field splitting effect \cite{gatteschi_molecular_2006,rudowicz_2015}.

In magnetic units with an intricate
bridging structure 
\cite{mccleverty_role_1998,muller_2000,vignesh_quenching_2017} the
mechanism of electrons' exchange may involve a
large number of intermediate atoms of different sorts. This imply that the correlations between electrons shared by
individual metal centers contribute to the
intrinsic magnetic features of these compounds. Thus, the constituent electrons are 
delocalized to a large extent
occupying molecular orbitals.
Depending on the spatial position of each metal center, the chemical 
composition of the bridges
connecting two adjacent centers, the change in temperature and
the action of external fields,
the electrons' distribution over the bridges may not be uniform.
In other words, these compounds may allow multiple energetically favorable spin-orbital configurations 
giving rise to multiple independent magnetic excitations that are not related to local anisotropy 
effects nor to the existence of electronic bands.
Studying the magnetic properties of these isolated molecular magnets
requires the application of advanced methods,
such as the multi-configurational self-consistent
field. However, 
in contrast to single molecular magnets with trivial bridging
structure and
due to the lack of sufficient number of model parameters, the outcome of such advanced
methods cannot be accounted for by
conventional spin models.
Neither the isotropic spin bilinear form nor higher-order
spin interaction terms can
account for all possible magnetic transitions and dependencies arising due to the delocalization of electrons.
Moreover, neither the zero field splitting
nor the exchange anisotropy terms are physically relevant to
such process.
Therefore, any attempt to gain insights into the existing exchange mechanisms
by combining different conventional Hamiltonians will be inadequate.

The present paper develops a theoretical approach, used in our
previous papers 
\cite{georgiev_mexchange_2019,georgiev_trimer_2019,georgiev_epjb_2019,georgiev_magnetization_2020},
to determine the
magnetic properties of isolated molecular magnets with non-trivial
bridging structure.
The named approach combines two widely
used methods, namely the molecular orbital theory and
the multi-configuration 
self-consistent field formalism resulting in a post-Hartree-Fock scheme 
for constructing the relevant energy spectrum. The main features of
the ensuing energy sequence
accounted for by constructing an effective spin-like microscopic model with
a few adjustable running parameters.
The named spin-like model was successfully applied to the
identification of the magnetic spectra 
of the spin-half trimeric compounds 
family A$_3$Cu$_3$(PO$_4$)$_4$ A=(Ca, Sr, Pb) \cite{georgiev_trimer_2019}
and the spin-one tetrameric molecular magnet Ni$_4$Mo$_{12}$ \cite{georgiev_epjb_2019}.
A good agreement with inelastic neutron scattering
experimental data was demonstrated.
Furthermore, the ensuing model adequately reproduces the magnetization and low-field
susceptibility measurements performed on the nickel tetramer Ni$_4$Mo$_{12}$
\cite{georgiev_magnetization_2020}.
In contrast to conventional methods that require distinct spin
Hamiltonians in order to explain the main features of each individual experiment, 
the constructed model includes a unique set
of parameters for any
experimental measurements.

The rest of the paper is organized as follows.
In Sec. \ref{sec:MO} we outline the basic principles of 
the molecular orbital theory and define the molecular orbitals to be further applied.
In Sec. \ref{sec:PHF} we represent the mathematical formalism aiming
to unveil the
mechanisms of exchange and to quantify the magnetic properties in molecular magnets with non-trivial 
bridging structure. Thus, we introduce the total
Hamiltonian and construct the variational functions. 
Moreover, we derive the corresponding
effective matrices and compute the generalized energy spectrum. 
In Sec. \ref{sec:SSH} we present the spin-like Hamiltonian accounting for the main
features predicted by the applied variational method. Further, we discuss technical 
aspects of the relevant spin-like operators. 
A primer for the application of the proposed method and spin-like Hamiltonian is 
given in Sec. \ref{sec:Dimer}.
Section \ref{sec:Conclusion} contains a summary of the introduced method and results.


\section{Molecular orbital theory}\label{sec:MO}

\subsection{First principles}\label{sec:view}

Nanomagnets possess unique properties
with respect to their size and spatial symmetry and
hence the corresponding chemical bonds length, angles and strength.
These characteristics are a consequence of the overlap between pure or 
hybridized orbital states of the valence electrons interacting with each 
other and with the nuclei of the constituent atoms. The electrons distribute 
themselves in such a way to minimize the molecule's total energy. 
Thus, the corresponding spatial distribution depends on the
constituent electrons' spin and therefore gives rise to a variety of
magnetic effects.

There are two main theories describing chemical bonds in materials:
the valence bond theory 
\cite{epiotis_unified_1983,cooper_valence_2002,shaik_chemists_2007}
and the molecular orbital theory
\cite{pople_approximate_1970,fleming_molecular_2009,evarestov_quantum_2012,albright_orbital_2013}. 
Both theories complement each other and successfully describe the quantum origin of chemical
bonding. By analogy to the constructive and destructive interference when combining 
wave functions and with respect to the choice of internuclear axis the
bonds are of two sorts: $\sigma$ and $\pi$ types. The strongest bonds are of  
$\sigma$ type and the associated electrons distributions are
energetically more favorable.

The valence bond theory includes in the bonding processes only
unpaired orbital
electrons from the most outer shells and does not account for the contribution of the remaining 
paired valence electrons. Moreover, in explaining the chemical bonds, this theory 
regards the electrons as localized in between
the constituent nuclei in pairs. 
Although, it is essential for the description of a multiple atomic
molecule, it raises difficulties in explaining the behavior of paramagnetic molecules.
Since magnetic exchange phenomena originate from the delocalization of
electrons, such localized approach 
hinders the derivation of the 
full energy spectra of the exchange interactions. 
Thus, an alternative approach
involving delocalized electrons may prove to be very
useful. It lies on a basic principle of the molecular orbital theory,
where the electrons are considered as spatially distributed over the entire molecule. Similar to the electron 
configurations in atoms, in molecular orbital theory electrons occupy 
molecular orbitals.
Along these lines, the ligand field theory combines both the
molecular orbital theory and the crystal field one
to explain the electronic structure and spin state of coordination complexes \cite{hoffmann_widely_2016}
and to study the magnetic properties of larger magnetic clusters
\cite{fraser_2017,loose_2008,yoon_electronic_2007}.

One way of constructing molecular orbitals is representing each one as a linear 
combination of atomic orbitals.
Depending on the problem under consideration one may apply different approximations
of atomic orbitals to minimize the energy, such as Hydrogen, Slater or
Gaussian types \cite{magnasco_2007,magnasco_2013}.
With respect to the type of bonds included in the construction of molecular orbitals,
i.e. $\sigma$ and $\pi$ types, these are divided in three groups:
bonding, antibonding and non-bonding. They are arranged in accordance to their energies. 
Lower in energy and hence most favorable are the bonding molecular orbitals, with energy
less than that of the constituent atomic orbitals. 
Contrary, less energetically favorable are the antibonding molecular orbitals.  
Non-bonding orbitals have no contribution to the bond processes and have the same energy
as the atomic orbitals they are constructed of.
Moreover molecular orbitals are classified as core, active and virtual.
Core orbitals are occupied by two electrons, also known as doubly or fully occupied
by virtue of the Pauli exclusion principle. The second class, active molecular orbitals are those
occupied by a single electron. They are also called half-filled. 
The virtual class represents the non-occupied molecular orbitals.

\subsection{Molecular orbitals}\label{sec:orbitals}

In constructing the $i$-th electron molecular orbital state 
we assume that the corresponding atomic state taken with respect to
the $\eta$-th constituent atom
within the molecule is given by
\begin{equation}\label{eq:AtomEigenstates}
	\psi^{\eta}_{\mu_{i,\eta},m_i}(\mathbf{r}_i)
	\equiv
	\psi^{\eta}_{\mu_{i,\eta}}(\mathbf{r}_i)
	\lvert m_i\rangle,
\end{equation}
where $\mu_{i,\eta}$ labels the shell and sub-shell that the $i$-th
electron occupies on the $\eta$-th atom, 
$\mathbf{r}_i=(x_i,y_i,z_i)$ and $m_i=\pm\tfrac{1}{2}$ are the $i$-th electron's coordinates 
and spin-magnetic quantum number, respectively. 
For convenience the variational parameters in
\eqref{eq:AtomEigenstates} are omitted. 
We would like to point out that in general the label $\mu_{i,\eta}$ may indicate either pure or 
hybridized atomic orbitals.
For example, if the $i$-th electron occupies some of the three $2p$
orbitals in the $\eta$-th atom, 
then $\mu_{i,\eta}\in\{2p_x,2p_y,2p_z\}$ and the 
state in \eqref{eq:AtomEigenstates} will be rewritten as 
$\psi^{\eta}_{2p_{\alpha},m_i}(\mathbf{r}_i)$, where
$\alpha\in\mathbb{K}$. 
When the occupied orbital is a hybridization between $2s$ and $2p_z$ 
orbitals one has $\mu_{i,\eta}\in\{2sp^{\pm}_z\}$. 
Moreover, in the presence of an external magnetic field all orbitals are considered as
gauge-invariant.

For all $\eta$ and $i$ the functions in \eqref{eq:AtomEigenstates} are orthogonal 
and normalized, i.e.,
\begin{equation}\label{eq:NormAtomState}
	\int
	\bar{\psi}^{\eta}_{\mu^{}_{i,\eta},m^{}_i}(\mathbf{r}_i)
	\psi^{\eta}_{\mu'_{i,\eta},m'_i}(\mathbf{r}_i)\mathrm{d}\mathbf{r}_i
	= \delta_{\mu^{}_{i,\eta}\mu'_{i,\eta}}\delta_{m^{}_i m'_i}.	
\end{equation}
Further, the overlap integral between any two arbitrary atomic states 
of the $i$-th electron residing on different atoms satisfies
\begin{equation}\label{eq:AtomOverlap}
	0\le\int
	\bar{\psi}^{\eta^{}}_{\mu^{}_{i,\eta},m_i}(\mathbf{r}^{}_i)
	\psi^{\eta'}_{\mu'_{i,\eta'},m_i}(\mathbf{r}_i)
	\mathrm{d}\mathbf{r}_i<1.
\end{equation}

With respect to the state functions \eqref{eq:AtomEigenstates} and the Born--Oppenheimer approximation, 
we represent the $n$-th molecular orbital state of the $i$-th electron by
\begin{equation}\label{eq:MOrbitals}
	\phi_{n,m_i}(\mathbf{r}_i)=
	\sum_{\eta} c^{\eta}_{n}
	\psi^{\eta}_{\mu_{i,\eta},m_i}(\mathbf{r}_i),
\end{equation}
where the molecular orbital number $n$ is determined according to the symmetry of all 
atomic shells, sub-shells 
and the type of bonds, i.e. $\sigma$ and $\pi$ bonds. 
The coefficients $c^{\eta}_{n}$ are functions of the overlap integrals between 
the atomic orbitals, see \eqref{eq:AtomOverlap}. 
Employing the bra-ket notation in \eqref{eq:AtomEigenstates} 
the molecular orbital state representation \eqref{eq:MOrbitals} reads
\begin{equation}\label{eq:MObra-ket}
	\phi_{n,m_i}(\mathbf{r}_i)\equiv\phi_{n}(\mathbf{r}_i)\lvert m_i\rangle
	=\prod_{{\alpha\in\mathbb{K}}}\phi_{n}(\alpha_i)\lvert
	m_i\rangle.
\end{equation}
For all $i$ the molecular orbitals given in \eqref{eq:MOrbitals} are orthogonal,
\begin{equation}\label{eq:OrthogonalOrbitals}
	\int
	\bar{\phi}_{n^{},m^{}_i}(\mathbf{r}_i)
	\phi_{n',m'_i}(\mathbf{r}_i) 
	\mathrm{d}\mathbf{r}_i = 
	\delta_{nn'}\delta_{m^{}_im'_i}.
\end{equation}

The arrangement of molecular orbitals in the energy diagram follows the value of $n$.
According to the \textit{Aufbau} principle 
their numeration is 
such that the lowest value of $n$ in \eqref{eq:MOrbitals} corresponds to the most 
energetically favorable molecular orbital.
Thus, the first occupied molecular orbital should
correspond the lowest level of the spectrum.


\section{Towards a post-Hartree-Fock method}\label{sec:PHF}

Consider a molecular magnet composed of an arbitrary number of atoms.  
We have a multi particle non-relativistic system obeying the adiabatic 
approximation.
Since the intrinsic electromagnetic field of the
molecule, due to the presence 
of electrons and nuclei, includes static electric and magnetic components,
henceforth we take into account only the dipole terms of the corresponding
electron-electron and electron-nuclei interactions. We neglect the effect of all 
orbital-orbital, spin-spin and spin-orbital interactions between the electrons and atomic nuclei.
The nuclei-nuclei potentials are also neglected, since
for the considered system 
they do not contribute to the magnetic properties. 
Moreover the magnitude of
the total magnetic vector potential is constant.

\subsection{Electron's generalized momentum}\label{sec:ElMomenta}

Let $\hat{\mathbf{p}}_{i}=(p^\alpha_i)_{\alpha\in\mathbb{K}}$ be the $i$-th electron's
momentum operator with coordinates
$\mathbf{r}_i=(\alpha_i)_{\alpha\in\mathbb{K}}$, 
$\hat{\boldsymbol{s}}_i=(\hat{s}^\alpha_i)_{\alpha\in\mathbb{K}}$ and 
$\hat{\boldsymbol{l}}_i=(\hat{l}^\alpha_i)_{\alpha\in\mathbb{K}}$
are, respectively, the corresponding spin and angular momentum 
operators, where $\mathbb{K}=\{x,y,z\}$ and $i=1,\ldots,N$.
Furthermore, let 
$\hat{\boldsymbol{\mu}}_{l_i}=-\mu_{\mathrm{B}}\hat{\boldsymbol{l}}_i$ 
and 
$\hat{\boldsymbol{\mu}}_{s_i}=-g_s\mu_{\mathrm{B}}\hat{\boldsymbol{s}}_i$ 
be the orbital and spin magnetic moments of the $i$-th electron, where 
$\mu_{\mathrm{B}}=e\hbar/2m_e$ is the Bohr magneton with 
$m_e$ denoting electron's rest mass.
The action of an external magnetic and the cluster's intrinsic magnetic fields into the 
$i$-th electron momentum is accounted for by means of the generalized momentum operator
$\hat{\mathbf{P}}_{i}=\hat{\mathbf{p}}_{i}-e\hat{\mathbf{A}}(\mathbf{r}_i)$,
where $\hat{\mathbf{A}}(\mathbf{r}_i)=(\hat{A}_\alpha(\mathbf{r}_i))_{\alpha\in\mathbb{K}}$ 
is the operator related to the total magnetic vector potential at a point with coordinates 
$\mathbf{r}_i$. We have
\begin{equation}\label{eq:VectorPotential}
	\hat{\mathbf{A}}(\mathbf{r}_i)=\hat{\mathbf{A}}_{ext}(\mathbf{r}_i)+ 
	\sum_{j|_{j\ne i}}\hat{\mathbf{A}}(\mathbf{r}_{ij}),
\end{equation}
where $\mathbf{r}_{ij}$ is the vector of the distance between $i$-th and 
$j$-th electrons.
In equation \eqref{eq:VectorPotential}, 
$\hat{\mathbf{A}}_{ext}=\big(\hat{A}^\alpha_{ext}(\mathbf{r}_i)\big)_{\alpha\in\mathbb{K}}$,
for $\hat{\mathbf{A}}_{ext}\equiv\mathbf{A}_{ext}$, 
represents the magnetic vector potential associated to the externally 
applied magnetic field $\mathbf{B}=(B_\alpha)_{\alpha\in\mathbb{K}}$ and 
$\hat{\mathbf{A}}(\mathbf{r}_{ij})=
\big(\hat{A}_\alpha(\mathbf{r}_{ij})\big)_{\alpha\in\mathbb{K}}$
the magnetic vector potential related to the
orbital angular momentum operator of the $j$-th electron.
In other words, under the considered approximations, we have
\begin{subequations}\label{eq:ExplicitVectorPotential}
\begin{equation}\label{eq:ExplicitVectorPotential_a}
	\mathbf{A}_{ext}(\mathbf{r})=\frac{1}{2}\mathbf{B}\times\mathbf{r}
\end{equation}
and
\begin{equation}\label{eq:ExplicitVectorPotential_b}
	\hat{\mathbf{A}}(\mathbf{r}_{ij})=
	\frac{\mu_0\mu_{\mathrm{B}}}{4\pi^2\langle r_{ij}\rangle^{3}}
	\mathbf{r}_{ij}\times(\hat{\boldsymbol{l}}_j+g_s\hat{\boldsymbol{s}}_j),
\end{equation}
\end{subequations}
where $\mu_0$ is the magnetic permeability in vacuum, 
$r_{ij}=|\mathbf{r}_i-\mathbf{r}_j|$ is the distance between $i$-th 
and $j$-th electrons and $\mathbf{r}\in\{\mathbf{r}_1,\ldots,\mathbf{r}_N\}$.
The average $\langle r_{ij}\rangle$ is taken with respect to the 
molecular orbitals $i$-th and $j$-th electrons occupy. 

Therefore, since for all $\alpha$ and $i$ the operators $\hat{A}_\alpha(\mathbf{r}_i)$ 
and $\hat{p}^\alpha_i$ commute, with respect to equations \eqref{eq:VectorPotential} 
and \eqref{eq:ExplicitVectorPotential} the generalized momentum operator reads
\begin{subequations}\label{eq:GeneralMomenum} 
\begin{align}
	\hat{\mathbf{P}}^2_{i} = & \ 
	\hat{\mathbf{p}}^2_{i}+2m_e \!\!
	\sum_{j|_{j\ne i}}\hat{K}^l_{ij}+ 2m_e \!\!
	\sum_{j|_{j\ne i}}\hat{K}^s_{ij}
\nonumber \\
	& -
	2m_e\mu_{\mathrm{B}}\mathbf{B}\cdot\hat{\boldsymbol{l}}_i+
	\frac{e^2}{4}\mathbf{B}^2\mathbf{r}^2_{i},
\end{align}
where
\begin{align}\label{eq:K_a}
\hat{K}^l_{ij}
= &-\frac{\mu_0\mu^2_{\mathrm{B}}}{2\pi^2\langle r_{ij}\rangle^{3}}
\hat{\boldsymbol{l}}_i\cdot\hat{\boldsymbol{l}}_j
	\nonumber \\ 
& + \frac{e^2\mu_0\mu_{\mathrm{B}}}{8m_e\pi^2\langle r_{ij}\rangle^{3}}
\bigg[ 
\left(\mathbf{r}_{ij}\cdot\mathbf{B}\right) 
\left(\mathbf{r}_{i}\cdot\hat{\boldsymbol{l}}_j\right)
-
\left(\mathbf{r}_{i}\cdot\mathbf{r}_{ij}\right)
\left(\mathbf{B}\cdot\hat{\boldsymbol{l}}_j\right)
\bigg],
\end{align}
\begin{align}\label{eq:K_b}
\hat{K}^s_{ij}
=&-\frac{\mu_0\mu^2_{\mathrm{B}}g_s}{2\pi^2\langle r_{ij}\rangle^{3}}
\hat{\boldsymbol{l}}_i\cdot\hat{\boldsymbol{s}}_j
\nonumber \\ 
&+ \frac{e^2\mu_0\mu_{\mathrm{B}}g_s}{8m_e\pi^2\langle r_{ij}\rangle^{3}}
\bigg[ 
\left(\mathbf{r}_{ij}\cdot\mathbf{B}\right) 
\left(\mathbf{r}_{i}\cdot\hat{\boldsymbol{s}}_j\right)
-
\left(\mathbf{r}_{i}\cdot\mathbf{r}_{ij}\right)
\left(\mathbf{B}\cdot\hat{\boldsymbol{s}}_j\right)
\bigg].
\end{align}
\end{subequations}
For further details on the representation of the $i$-th electron's general momentum 
operator in \eqref{eq:GeneralMomenum}, the interested reader
may consult Appendix \ref{sec:AppGenMom}
and Refs. 
\cite{blundell_magnetism_2001,getzlaff_fundamentals_2008}.

\subsection{The canonical Hamiltonian}

According to all assumptions and definitions in Sec.
\ref{sec:ElMomenta},
the Hamiltonian accounting for all interactions in the considered system reads
\begin{equation}\label{eq:CoulombHamilton}
	\hat{H}=\sum_{i}\frac{{\hat{\mathbf{P}}^2_{i}}}{2m_e}+
	\sum_{\eta,i}\hat{U}_{\eta i}+\frac12\sum_{i\ne j}\hat{R}_{ij}
	-g_s\mu_{\mathrm{B}}\sum_{i}\mathbf{B}\cdot\hat{\boldsymbol{s}}_i,
\end{equation}
where the potential energy operator 
$\hat{U}_{\eta i}\equiv U(r_{\eta i})$ accounts for the interaction of the $i$-th electron 
with the $\eta$-th nucleus, separated by the distance 
$r_{\eta i} = |\mathbf{r}_{i}-\mathbf{R}_{\eta}|$, with
$\mathbf{R}_{\eta}$ representing the coordinates of the $\eta$-th nucleus.
The operator $\hat{R}_{ij}\equiv R(r_{ij})$ is related to the repulsion potential
between $i$-th and $j$-th electrons over the distance 
$r_{ij}=|\mathbf{r}_i-\mathbf{r}_j|$.

Taking into account equations
\eqref{eq:GeneralMomenum}, Hamiltonian
\eqref{eq:CoulombHamilton} can be represented as 
the sum of three terms, i.e. 
\begin{equation}\label{eq:CoulombHamilton_2}
	\hat{H}=\hat{H}_r+\hat{H}_l+\hat{H}_s,
\end{equation}
where the spin-independent part of the total Hamiltonian is given by
\begin{subequations}\label{eq:PartialHamiltonians}
\begin{align}\label{eq:SpatialHamilton}
	\hat{H}_r 
	= & 
	\sum_{i}\frac{{\hat{\mathbf{p}}^2_{i}}}{2m_e}+
	\sum_{\eta,i}\hat{U}_{\eta i}+
	\sum_{i\ne j}
	\frac12\hat{R}_{ij}-
	\sum_i\frac{e^2}{8m_e}\mathbf{B}^2\mathbf{r}^2_{i},
\end{align}
the orbital term reads
\begin{align}\label{eq:OrbitalHamiltonian}
\hat{H}_l
= \sum_{i\ne j}\hat{K}^l_{ij}-
\mu_{\mathrm{B}}\sum_{i}\mathbf{B}\cdot\hat{\boldsymbol{l}}_i
\end{align}
and the spin one is
\begin{align}\label{eq:SpinHamilton}
	\hat{H}_s
	= \sum_{i\ne j}\hat{K}^s_{ij} -
	g_s\mu_{\mathrm{B}}\sum_{i}\mathbf{B}\cdot\hat{\boldsymbol{s}}_i.
\end{align}
\end{subequations}

For $|\mathbf{B}|>0$, the orbital bilinear form in \eqref{eq:K_a} and the 
spin-orbit form in \eqref{eq:K_b} are negligibly small
compared to the other terms. 
As a result, the expressions of the respective Hamiltonians given in \eqref{eq:OrbitalHamiltonian} and 
\eqref{eq:SpinHamilton} simplify. Thus,
We have
\begin{equation}\label{eq:OrbitalHamilton_g}
\hat{H}_l= -\mu_{\mathrm{B}}B \sum^{N}_{i=1}
\boldsymbol{g}_i\cdot\hat{\boldsymbol{l}}_i
\end{equation}
and
\begin{equation}\label{eq:SpinHamilton_g}
\hat{H}_s= -g_s\mu_{\mathrm{B}}B \sum^{N}_{i=1}
\boldsymbol{g}_{i}\cdot\hat{\boldsymbol{s}}_i,
\end{equation}
where $\mathbf{B}=\mathbf{n}B$, $\mathbf{n}=(n_\alpha)_{{\alpha\in\mathbb{K}}}$
is a unit vector and the $\boldsymbol{g}$ factor reads
\begin{equation}\label{eq:G-Factor}
	\boldsymbol{g}_i=
	\left( \mathbf{n}-
	\sum^{}_{j|_{j\ne i}}\frac{e^2\mu_0}{8m_e\pi^2\langle r_{ij}\rangle^{3}}
	\bigg[ 
	\mathbf{r}_{j}\left(\mathbf{r}_{ij}\cdot\mathbf{n}\right)-
	\mathbf{n}\left(\mathbf{r}_{j}\cdot\mathbf{r}_{ij}\right)
	\bigg]
	\right),
\end{equation}
with $\boldsymbol{g}_i=(g^\alpha_i)_{\alpha\in\mathbb{K}}$.

\subsection{State functions}\label{sec:SF}

Studying the magnetic behavior of a molecule with complex bridging structures
we apply the molecular orbital theory and construct the variational state function
within the multi-configurational self-consistent field method.
Thus, selecting $\alpha\in\mathbb{K}$ axis as a quantization axis,
we describe an arbitrary spin configuration of 
$N$ number of electrons, occupying $n$ molecular orbitals
given in \eqref{eq:MOrbitals}, by
\begin{align}\label{eq:MoleculeBasisState}
	\Psi^{1,\ldots,n-1,n}_{m_1,\ldots, m_N}(\mathbf{r}_1,\ldots,\mathbf{r}_N)=
	&
	\sum_{P_{\mathbf{r}_1,\ldots, \mathbf{r}_N}} 
	c_{\mathbf{r}_1,\ldots,\mathbf{r}_N}
	\phi_{1,m_1}(\mathbf{r}_1)\phi_{1,m_2}(\mathbf{r}_2)
\nonumber \\ 
&	
	\cdots\phi_{{n-1},m_{N-1}}(\mathbf{r}_{N-1})
	\phi_{n,m_N}(\mathbf{r}_N),
\end{align} 
where $n-1$ and $n$ are the highest in energy active molecular orbitals,
the sum runs over all permutations on the set of
coordinates $\mathbf{r}_1,\ldots, \mathbf{r}_{N}$, the coefficients
$c_{\mathbf{r}_1, \ldots, \mathbf{r}_N}$ account for the orthonormalization and antisymmetry
of the state function and the set $\{m_1,\ldots, m_N\}$
identifies the spin configuration
according to the selected quantization axis.
Each function in \eqref{eq:MoleculeBasisState} corresponds to a single Slater determinant.

In general, all $N$ electrons are involved in the exchange process. However,
as the core molecular orbitals are permanently occupied by two electrons 
the total spin multiplet scheme
of the system is generated with respect to the number of all active and post-active orbitals.
Furthermore, since a triplet-singlet transition of an electron spin pair requires two active orbitals
each individual spin multiplet relates to certain pairs of such orbitals.
Therefore, let $\tau\in\{(n-i,n-j),\ldots,(n-1,n)\}$ indicate all pairs or 
a tuple of pairs of active orbitals, with $i\ne j=2,3\ldots,n-(\bar{n}+1)$, where $\bar{n}$ is
the number of the highest in energy core molecular orbital.
Then, the generalized state function representing a spin multiplet related to the pair of active orbitals
$\tau=(n-1,n)$ is given by the following superposition of the basis states in \eqref{eq:MoleculeBasisState}
\begin{align}\label{eq:OrbitalSymmetrizedBasisStates}
	\Psi^{\tau,v_{\tau,s_\tau,s}}_{s_\tau,s,m}
	(\mathbf{r}_1,\ldots,\mathbf{r}_N)=
	&
	\sum_{m_1=-s_1}^{s_1}\! \ldots \! \sum_{m_N=-s_N}^{s_N}
	c^{s_\tau,s,m}_{m_1, \ldots,m_N}
\nonumber \\
	&\times\!\!\!
	\sum_{n-1,n}{ c^{v_{\tau,s_\tau,s}}_{n-1,n}}
	\Psi^{1,\ldots,n-1,n}_{m_1,\ldots, m_N}(\mathbf{r}_1,\ldots,\mathbf{r}_N),
\end{align}
where 
$c^{s_\tau,s,m}_{m_1, \ldots,m_N}$ are the Clebsch-Gordan coefficients,
the scalars { $c^{v_{\tau,s_\tau,s}}_{n-i,n}$} symmetrize the state function
with respect to all energetically equivalent orbitals within the considered pair
or a tuple of pairs, { with $v_{\tau,s_\tau,s}\in\mathbb{N}$ indicating all linearly independent combinations,} 
$s$ and $m$ denote the system's total spin quantum numbers
and $s_\tau\equiv s_{n-1,n}$ the local spin quantum number related to both active orbitals.
We would like to point out that in both \eqref{eq:MoleculeBasisState} and 
\eqref{eq:OrbitalSymmetrizedBasisStates} there is a sum running over the 
permutation of all electrons. As a result, the quantum number
$s_{n-1,n}$ will represent the spin quantum number of each electron pair
occupying the orbitals $n-1$ and $n$.
Furthermore, the array of subscripts $s_\tau,s,m$ is a
shorthand notation of the
resulting spin multiplet represented by all good spin quantum numbers
$\{s_{n-i,n-j},\ldots, s_{n-1,n},s,m\}$. For the sake of clarity 
we use only spin quantum number relevant to all active orbitals.
For example, if the configuration includes four active orbitals, 
with $\tau=\{(n-3,n-2),(n-1,n)\}$, 
then the set $s_\tau,s,m$ should be rewritten as $s_{n-3,n-2},s_{n-1,n},s,m$, see also
Appendix \ref{sec:AppFourElectrons}.

The number of all sets of molecular orbitals involved in the exchange processes $\tau$ 
increases with the number of all intermediate bridges.
Respectively, the number of all state functions in \eqref{eq:OrbitalSymmetrizedBasisStates} will depend
on all distinct electrons' distributions and 
hence spin-orbital configurations indicated by the index $\tau$.  
The electrons' correlations for any 
such distribution are taken into account with respect to the effective 
Hamiltonian $\hat{H}_{\mathrm{eff}}=\hat{H}^r_{\mathrm{eff}}+\hat{H}^l_{\mathrm{eff}}+\hat{H}^s_{\mathrm{eff}}$
to the initial Hamiltonian given in \eqref{eq:CoulombHamilton_2}, that
accounts for the energy levels
degeneracy with regard to the spatial and
spin components of the considered state functions.
The effective semi-spin states, corresponding to these given in
\eqref{eq:OrbitalSymmetrizedBasisStates},
read { $\lvert\Psi^{\tau,v_{\tau,s_\tau,s}}_{s_\tau,s,m}\rangle$}.
Hence, constructing the total effective matrix, we have
\begin{align}\label{eq:GeneralEffHam}
	\Big\langle\Psi^{\tau,v_{\tau,s_\tau,s}}_{s_\tau,s,m}\Big\rvert	
	\hat{H}_{\mathrm{eff}}
	\Big\lvert\Psi^{\tau,v_{\tau,s_\tau,s}}_{s_{\tau},s,m}\Big\rangle\equiv
	&
	\int\!\cdots\!\int 
	\bar{\Psi}^{\tau,v_{\tau,s_\tau,s}}_{s_\tau,s,m}(\mathbf{r}_1,\ldots,\mathbf{r}_N)
	\hat{H}
\nonumber \\
	& 
	\times\Psi^{\tau,v_{\tau,s_\tau,s}}_{s_{\tau},s,m}(\mathbf{r}_1,\ldots,\mathbf{r}_N) 
	\mathrm{d}\mathbf{r}_1\cdots\mathrm{d}\mathbf{r}_N.
\end{align}
Since Hamiltonian \eqref{eq:CoulombHamilton} accounts for only
pair interactions and the terms $\hat{H}_r$ and $\hat{H}_l$ act only
on the spatial component of 
\eqref{eq:OrbitalSymmetrizedBasisStates}, the effective matrices associated to 
$\hat{H}^r_{\mathrm{eff}}$ and $\hat{H}^l_{\mathrm{eff}}$ will have off-diagonal terms only 
with respect to $v_{\tau,s_\tau,s}$. 
Accordingly, the diagonalization of the total matrix related to $\hat{H}_{\mathrm{eff}}$ 
will mix the elements of both matrices associated to $\hat{H}^r_{\mathrm{eff}}$
and $\hat{H}^l_{\mathrm{eff}}$. As a result, in diagonal form we have
$\hat{H}_{\mathrm{eff}}\to\hat{\mathrm{H}}^{r,l}_{\mathrm{eff}}+\hat{\mathrm{H}}^s_{\mathrm{eff}}$, such that
\begin{subequations}\label{eq:BeforeGenericStateFunction}
\begin{equation}\label{eq:BeforeGenericStateFunction_a}
	\hat{\mathrm{H}}^{r,l}_{\mathrm{eff}}
	\big\lvert\Omega^{\tau,n_{\tau,s_\tau,s}}_{s_\tau,s,m}\big\rangle=
	E^{\mathrm{f},(n_{\tau,s_\tau,s})}_{\tau,s_\tau,s,m}
	\big\lvert\Omega^{\tau,n_{\tau,s_\tau,s}}_{s_\tau,s,m}\big\rangle,
\end{equation}
\begin{equation}\label{eq:BeforeGenericStateFunction_b}
	\hat{\mathrm{H}}^s_{\mathrm{eff}}
	\big\lvert\Omega^{\tau,n_{\tau,s_\tau,s}}_{s_\tau,s,m}\big\rangle=
	\mathcal{E}^{(n_{\tau,s_\tau,s})}_{\tau,s_\tau,s,m}
	\big\lvert\Omega^{\tau,n_{\tau,s_\tau,s}}_{s_\tau,s,m}\big\rangle,
\end{equation}
\end{subequations}
where 
the index $n_{\tau,s_\tau,s}\in\mathbb{N}$ runs
over the corresponding 
subset of all eigenstates given by
\begin{equation}\label{eq:StatesAfterDiagonalization}
\big\lvert\Omega^{\tau,n_{\tau,s_\tau,s}}_{s_\tau,s,m}\big\rangle=
{ \sum_{v_{\tau,s_\tau,s}}c^{v_{\tau,s_\tau,s}}_{n_{\tau,s_\tau,s}}
\big\lvert\Psi^{\tau,v_{\tau,s_\tau,s}}_{s_\tau,s,m}\big\rangle}.
\end{equation}
and the superscript ``f'' indicates that in the case $B\ne0$ the eigenvalues
\eqref{eq:BeforeGenericStateFunction_a} are functions of the
externally applied magnetic field. 
Notice that after diagonalization the initial Hilbert space relevant 
to \eqref{eq:OrbitalSymmetrizedBasisStates} is reduced and the
eigenstates \eqref{eq:BeforeGenericStateFunction} take into account only the 
spins related to the active molecular orbitals.

Due to the existence of different spin-orbital configurations indicated by the { indexes $\tau$
and $n_{\tau,s_\tau,s}$}, 
for a given total spin $s$ the basis set of all functions in \eqref{eq:StatesAfterDiagonalization}
includes an array of spin multiplets each distinguished by the quantum numbers $s_\tau$. 
Therefore, in constructing the final effective spin space
we incorporate the contributions of all individual spin multiplets,
{ $(\ldots,s_\tau,s,m),(\ldots,s'_\tau,s,m)\ldots,(\ldots,s'_{\tau'},s,m)$,}
by introducing the following superposition
\begin{equation}\label{eq:GenericStateFunction}
	\lvert s,m\rangle=\sum_{[\ldots,s_\tau]}\sum_{\tau,n_{\tau,s_\tau,s}}
	c^{(n_{\tau,s_\tau,s})}_{\tau,s_\tau,s,m}
	\big\lvert\Omega^{\tau,n_{\tau,s_\tau,s}}_{s_\tau,s,m}\big\rangle,
\end{equation}
where
the coefficients $c^{(n_{\tau,s_\tau,s})}_{\tau,s_\tau,s,m}$ are associated to the probability weights
accounting for the contribution of each energetically distinct spin-orbital 
configuration labeled by $\tau$ and $n_{\tau,s_\tau,s}$
for observing the system in a spin state characterized by the total quantum numbers $s$ and $m$.
At each total spin level, the first sum,
with $[\ldots,s_\tau]$ in \eqref{eq:GenericStateFunction}, runs only over 
the local spin quantum numbers related to the selected active molecular orbitals.
For example, if we consider four active orbitals with $\tau=\{(n-3,n-2),(n-1,n)\}$, 
then the sum will run over all values of $s_{n-3,n-2}$ and $s_{n-1,n}$, where
$|s_{n-3,n-2}+s_{n-1,n}|\geq s \geq|s_{n-3,n-2}-s_{n-1,n}|$.   
Therefore, in contrast to \eqref{eq:StatesAfterDiagonalization} the Hilbert space dimension 
associated to \eqref{eq:GenericStateFunction} is reduced to a large
extent rendering the
analytical computation with an
appropriately parametrized spin model more 
tractable.

\subsection{Key integrals}\label{sec:KI}

In the presence of
an external magnetic field the energy of each molecular orbital shifts
and hence the values of all exchange integrals alter.
Moreover, the indirect action of $\mathbf{B}$ on the
electrons' correlations is taken into account
by the field-dependent eigenvalues \eqref{eq:BeforeGenericStateFunction_a},
where the diamagnetic terms
\eqref{eq:SpatialHamilton} and the paramagnetic ones \eqref{eq:OrbitalHamiltonian}
contribute with different weights to these effective matrix's diagonal elements.

Within th convention, the post-coulomb integral referring to
the $(N-1)$-th and $N$-th electrons occupying the $n$-th molecular
orbital, with $n=N/2$, reads
\begin{subequations}\label{eq:GenericExchangeIntegrals}
	\begin{equation}\label{eq:GenericOneOrbitIntegral}
	U_{n}=
	\iint
	\bar{\phi}_{n}(\mathbf{r}_{N-1})
	\bar{\phi}_{n}(\mathbf{r}_{N})
	\hat{H}'_{r,l}
	\phi_{n}(\mathbf{r}_{N-1})
	\phi_{n}(\mathbf{r}_{N})
	\mathrm{d}\mathbf{r}_{N-1}
	\mathrm{d}\mathbf{r}_{N},
\end{equation}
where
\begin{align}\label{eq:LowOrbitalAverageEnergy}
	\hat{H}'_{r,l}
	=
	&
	\prod_{i,j}^{\tfrac{N}{2}-1}
	\int\!\!\cdots\!\!\int
	\bar{\phi}_j(\mathbf{r}_{2j-1})\bar{\phi}_j(\mathbf{r}_{2j})
	(\hat{H}_r+\hat{H}_l)
\nonumber \\
	& \times
	\phi_i(\mathbf{r}_{2i-1})\phi_i(\mathbf{r}_{2i})
	\mathrm{d}\mathbf{r}_{1}\cdots\mathrm{d}\mathbf{r}_{N-2},
\end{align}
is the Hamiltonian associated to the energy of $(N-1)$-th and 
$N$-th electrons interacting with the averaged field of all remaining electrons. 
Hamiltonian $\hat{H}'_{r,l}$ incorporates all correlations related to the
lowest in energy core molecular orbitals. 

The post-hopping integral corresponding to the same electrons, with 
$N$-th one hopping between $n$-th and $n'$-th molecular orbitals, is given by
\begin{equation}\label{eq:GenericOrbitHoppingIntegral}
	t_{nn'}=
	\iint
	\bar{\phi}_{n}(\mathbf{r}_{N-1})
	\bar{\phi}_{n}(\mathbf{r}_{N})
	\hat{H}'_{r,l}
	\phi_{n}(\mathbf{r}_{N-1})
	\phi_{n'}(\mathbf{r}_{N})
	\mathrm{d}\mathbf{r}_{N-1}
	\mathrm{d}\mathbf{r}_{N},
\end{equation}
where $n'=N/2-k$, for $k\in\mathbb{N}$. In particular, according to the
Hund's rule, $k=1$.
When both electrons occupy different orbitals, we have the second coulomb integral
\begin{equation}\label{eq:GenericTwoOrbitIntegral}
	V_{nn'}=
	\iint
	\bar{\phi}_{n}(\mathbf{r}_{N-1})
	\bar{\phi}_{n'}(\mathbf{r}_{N})
	\hat{H}'_{r,l}
	\phi_{n}(\mathbf{r}_{N-1})
	\phi_{n'}(\mathbf{r}_{N})
	\mathrm{d}\mathbf{r}_{N-1}
	\mathrm{d}\mathbf{r}_{N}. 
\end{equation}
	The direct exchange integral associated to the direct exchange of 
	$(N-1)$-th and $N$-th electrons between orbitals $n$ and $n'$ is
\begin{equation}\label{eq:GenericExchamgeOrbitIntegral}
	D_{nn'}=
	\iint
	\bar{\phi}_{n}(\mathbf{r}_{N-1})
	\bar{\phi}_{n'}(\mathbf{r}_{N})
	\hat{H}'_{r,l}
	\phi_{n'}(\mathbf{r}_{N-1})
	\phi_{n}(\mathbf{r}_{N})
	\mathrm{d}\mathbf{r}_{N-1}
	\mathrm{d}\mathbf{r}_{N}.
\end{equation}
\end{subequations}
After implicit summation over
all electrons and probability coefficients in
\eqref{eq:GeneralEffHam}, integrals
\eqref{eq:GenericExchangeIntegrals} determine the
eigenvalues \eqref{eq:BeforeGenericStateFunction_a} accordingly.

All eigenvalues in \eqref{eq:BeforeGenericStateFunction_b} account for the direct action
of the externally applied magnetic field, such that from \eqref{eq:SpinHamilton_g}
we have
\begin{equation}\label{eq:G-FactorAverage}
	\mathcal{E}^{(n_{\tau,s_\tau,s})}_{\tau,s_\tau,s,m}=
	-\mu_{\mathrm{B}}B \sum^{N}_{i=1}
	\langle g^{\alpha}_i\rangle^{(n_{\tau,s_\tau,s})}_{\tau,s_\tau,s}
	\langle\hat{s}^{\alpha}_i\rangle_{s_\tau,s,m},
\end{equation}
where the electron's $g_s$-factor is implicitly included in the corresponding average 
and with respect to the preselected quantization axis $\alpha$,
$\langle\hat{\boldsymbol{s}}^{}_i\rangle_{s_\tau,s,m}
\to\langle\hat{s}^{\alpha}_i\rangle_{s_\tau,s,m}$. 
Notice that depending on the existing spin-orbital configurations, after diagonalization, 
for all $\alpha$ the average value of the $g$-factor in \eqref{eq:G-FactorAverage}
may be a function of multiple average values 
$\langle g^{\alpha}_i\rangle^{(v_{\tau,s_\tau,s})}_{\tau,s_\tau,s}$ entering in 
\eqref{eq:GeneralEffHam}, with $v_{\tau,s_\tau,s}\in\mathbb{N}$.

\subsection{Energy spectrum}

In obtaining the energy sequence corresponding to the total effective spin space 
\eqref{eq:GenericStateFunction} we take into account the temperature at which 
measurements take place and imply the relation
\begin{equation}\label{eq:GenericEnergy}
	E^{\mathrm{f}}_{s,m}=
	\langle s,m\rvert\hat{\rho}_{s,m}
	\left(\hat{\mathrm{H}}^{r,l}_{\mathrm{eff}}+
	\hat{\mathrm{H}}^s_{\mathrm{eff}}\right) 
	\lvert s,m\rangle,
\end{equation}
where the operator
\begin{equation}\label{eq:DistOperator}
	\hat{\rho}_{s,m}=\sum_{[\ldots,s_\tau]}\sum_{\tau,n_{\tau,s_\tau,s}}
	\rho^{(n_{\tau,s_\tau,s})}_{\tau,s_\tau,s,m}
	\delta^{}_{\tau\tau'}\delta^{}_{s_\tau s'_\tau}
	\delta^{}_{n_{\tau,s_\tau,s}n'_{\tau,s_\tau,s}}
\end{equation}
accounts for the probability distribution 
for observing the system in a state with given
energy at given temperature. In terms of the Boltzmann
distribution, we have
\begin{equation}\label{eq:Boltzmann}
	\rho^{(n_{\tau,s_\tau,s})}_{\tau,s_\tau,s,m}=
	Z^{-1}e^{-\beta \varSigma^{\mathrm{f},(n_{\tau,s_\tau,s})}_{\tau,s_\tau,s,m}},
\end{equation}
where $\beta=1/\kappa_B T$ (with $\kappa_B$ the Boltzmann
constant), the partition function
\[
	Z=\sum_{\tau,n_{\tau,s_\tau,s}} \ \sum_{s_\tau,s,m}
	e^{-\beta \varSigma^{\mathrm{f},(n_{\tau,s_\tau,s})}_{\tau,s_\tau,s,m}}
\]
and 
\begin{equation}\label{eq:TotalEffEnergy}
	\varSigma^{\mathrm{f},(n_{\tau,s_\tau,s})}_{\tau,s_\tau,s,m}=
	E^{\mathrm{f},(n_{\tau,s_\tau,s})}_{\tau,s_\tau,s,m}+
	\mathcal{E}^{(n_{\tau,s_\tau,s})}_{\tau,s_\tau,s,m}.
\end{equation}
Now, accounting for \eqref{eq:GenericEnergy}, \eqref{eq:DistOperator} and
\eqref{eq:TotalEffEnergy}, for the effective energy spectrum we get
\begin{equation}\label{eq:GenericEnergyExplic}
	E^{\mathrm{f}}_{s,m}=\sum_{[\ldots,s_\tau]}\sum_{\tau,n_{\tau,s_\tau,s}}
	\left|c^{(n_{\tau,s_\tau,s})}_{\tau,s_\tau,s,m}\right|^2\rho^{(n_{\tau,s_\tau,s})}_{\tau,s_\tau,s,m}
	\varSigma^{\mathrm{f},(n_{\tau,s_\tau,s})}_{\tau,s_\tau,s,m}.
\end{equation}
Hence, an arbitrary energy gap in the obtained spectrum is given by
\begin{align}\label{eq:GenericEnergyTransit}
	\left|E^{\mathrm{f}}_{\Delta s,\Delta m}\right|=
&	
	\left| \sum_{\tau}
	\left( 
	\sum_{[\ldots,s_\tau]}\sum_{n_{\tau,s_\tau,s}}
	\left|c^{(n_{\tau,s_\tau,s})}_{\tau,s_\tau,s,m}\right|^2\rho^{(n_{\tau,s_\tau,s})}_{\tau,s_\tau,s,m}
	\varSigma^{\mathrm{f},(n_{\tau,s_\tau,s})}_{\tau,s_\tau,s,m}
	\right.
	\right.
\nonumber \\
&	\left.
	\left.
	-
	\sum_{[\ldots,s'_\tau]}\sum_{n_{\tau,s'_\tau,s'}}
	\left|c^{(n_{\tau,s'_\tau,s'})}_{\tau,s'_\tau,s',m'}\right|^2
	\rho^{(n_{\tau,s'_\tau,s'})}_{\tau,s'_\tau,s',m'}
	\varSigma^{\mathrm{f},(n_{\tau,s'_\tau,s'})}_{\tau,s'_\tau,s',m'} 
	\right) 
	\right|,
\end{align}
where $\Delta s=s-s'$ and $\Delta m=m-m'$. 
Within the last relation one has to bear in mind that the transitions 
between states characterized by a same local spin quantum number $s_\tau$ are
forbidden. 
Consequently, the energy of a magnetic transition associated to $\Delta s$ 
and $\Delta m$ will vary in accordance to each spin-orbital configuration
indicated by $\tau$ { and $n_{\tau,s_\tau,s}$}.

\section{The spin-sigma Hamiltonian}\label{sec:SSH}

A magnetic molecule with a particular size, symmetry and complexity
of the intermediate bridging structure among the magnetic
centers
may exhibits a unique set of energetically favorable electrons'
distributions.
Accordingly, being exposed to an external action it may respond uniquely in revealing 
its magnetic properties.
On the theoretical side, the relevant
magnetic features could be studied with the aid 
of the post-Hartree-Fock method discussed in Section \ref{sec:PHF}
and a spin Hamiltonian adequate to the computation of the
energy spectrum \eqref{eq:GenericEnergyExplic}.
Since the spectrum in \eqref{eq:GenericEnergyExplic} consists of two
independent components,
we have to rely on two different spin terms.
One Hamiltonian addressing the exchange interactions relevant to the eigenvalues 
\eqref{eq:BeforeGenericStateFunction_a} and
a Zeeman term incorporating the direct action of the external magnetic field according to
energy eigenvalues in \eqref{eq:BeforeGenericStateFunction_b}.

Within the considered method, due to the small number of effective
parameters,
spin
bilinear Hamiltonians, such as the Heisenberg model, 
may not account for all probable 
transitions in \eqref{eq:GenericEnergyTransit} at zero magnetic
field, $|\mathbf{B}|=0$. 
Respectively, the effective energy spectrum obtained with 
any such Hamiltonian will represent only a small fragment of
the full energy sequence relevant to 
the possible exchange processes.
Further, for $|\mathbf{B}|\ne0$ the conventional spin Zeeman term
cannot be used within the present method,
since the spectroscopic $g$-tensor is derived based on different
physical reasoning.
Under this circumstances, some splittings and broadening of the non-magnetic peaks in the 
corresponding magnetic spectrum may remain unexplained or
erroneously attributed 
to the magnetic anisotropy. Therefore, as it was argued in Sec. \ref{sec:intro}
any attempt to describe such effects by adding
different magneto-anisotropic terms or higher-order spin interaction
terms to the Heisenberg model will not be physically adequate.

To identify all possible
transitions described by \eqref{eq:GenericEnergyTransit},
we need at hand an appropriate bilinear spin
interaction form and a field term
with eigenstates relevant to \eqref{eq:GenericStateFunction}.
To this end, we propose the following Hamiltonians
\begin{subequations}\label{eq:TwoSpinSigmaHamiltonian}
	\begin{equation}\label{eq:SpinSigmaHamiltonian}
		\hat{\mathcal{H}}_{\sigma}=
		\sum\limits_{i \ne j}^{} J_{ij} 
		\hat{\boldsymbol{\sigma}}_i \cdot \hat{\mathbf{s}}_j
	\end{equation}
and
	\begin{equation}\label{eq:SigmaZeeman}
		\hat{\mathcal{H}}_{Z}=-\mu_{\mathrm{B}}B\sum_{i}
		\hat{S}^\alpha_i
	\end{equation}
\end{subequations}
where the couplings $J_{ij} = J_{ji}$ are effective exchange constants,
$\hat{\mathbf{s}}_i=(\hat{\mathrm{s}}^\alpha_i)_{\alpha\in\mathbb{K}}$, 
is the spin operator of the $i$-th effective magnetic center, for all $\alpha$ the sigma operator
$\hat{\boldsymbol{\sigma}}_i=(\hat{\sigma}^\alpha_i)_{\alpha\in\mathbb{K}}$ and
$\hat{S}^\alpha_i$ are functions of the $i$-th effective spin discussed hereafter.

In particular, Hamiltonian \eqref{eq:SpinSigmaHamiltonian} maps the energy gaps given in 
\eqref{eq:GenericEnergyTransit} only with respect to the terms 
$E^{\mathrm{f},(n_{\tau,s_\tau,s})}_{\tau,s_\tau,s,m}$.
The explicit representation and hence the 
physical meaning of all $J$-couplings differ from those 
entering the Heisenberg Hamiltonian. 
Moreover, as the energy terms $E^{\mathrm{f},(n_{\tau,s_\tau,s})}_{\tau,s_\tau,s,m}$ are field dependent 
the sigma operator aim to generate two sets of parameters.
The first one accounts for all 
transitions related to the probability amplitudes and distribution probabilities
\eqref{eq:GenericEnergyTransit} for $|\mathbf{B}|=0$. The second
set indicates the changes 
in all electrons' correlations due to the contribution of external magnetic field.

Hamiltonian \eqref{eq:SigmaZeeman} 
represents the Zeeman term and hence takes into account only the direct action of the 
external magnetic field related to all energy terms 
$\mathcal{E}^{(n_{\tau,s_\tau,s})}_{\tau,s_\tau,s,m}$
\eqref{eq:GenericEnergyExplic}.
Generally speaking, the spin-like operator
\eqref{eq:SigmaZeeman} is a function of the spin $g$-factor given in
\eqref{eq:G-FactorAverage} and the presence of only one component in
\eqref{eq:SigmaZeeman} is due to the preselected in Sec. \ref{sec:SF}
quantization axis.

Taking into account \eqref{eq:TwoSpinSigmaHamiltonian}, the energy spectrum given 
in \eqref{eq:GenericEnergyExplic} and the corresponding energy gaps 
\eqref{eq:GenericEnergyTransit},
for the total Hamiltonian describing an arbitrary single molecular magnet, we get 
\begin{equation}\label{eq:TSpinSigmaHamiltonian}
\hat{\mathcal{H}} = 
\sum\limits_{i \ne j}^{} J_{ij} 
\hat{\boldsymbol{\sigma}}_i \cdot \hat{\mathbf{s}}_j
-\mu_{\mathrm{B}}B\sum_{i}
\hat{S}^\alpha_i.
\end{equation}

Henceforth, we discuss in details the properties of the defined spin-like operators.


\subsection{Sigma operators}\label{sec:SS}

The components of the
$\boldsymbol{\sigma}$-operator are such that for all $i$,
\begin{equation}\label{eq:Sigma_i}
\hat{\sigma}^\alpha_i\lvert\ldots ,s_i,m_i,\ldots\rangle_{n_{s_i}}
=
a^{s_i}_{i,n_{s_i}} \hat{\mathrm{s}}^{\alpha_{\phantom{i}}}_i
\lvert\ldots ,s_i,m_i,\ldots\rangle_{n_{s_i}},
\end{equation}
where the used basis states are orthogonal with respect to $n_{s_i}\in\mathbb{N}$ and
$a^{s_i}_{i,n_{s_i}}$ are real parameters.
Moreover, the $\boldsymbol{\sigma}$
rising and lowering operators obey 
\begin{equation}\label{eq:Sigma_LR_i}
\hat{\sigma}^{\pm}_i\lvert\ldots ,s_i,m_i,\ldots\rangle_{n_{s_i}}
=
a^{s_i}_{i,n_{s_i}}\hat{\mathrm{s}}^{{\pm}_{\phantom{i}}}_i
\lvert\ldots ,s_i,m_i,\ldots\rangle_{n_{s_i}}.
\end{equation}
The square of $\hat{\boldsymbol{\sigma}}_i$ commutes only with its 
$z$ component
and according to \eqref{eq:Sigma_i} and \eqref{eq:Sigma_LR_i} 
it possesses the following eigenvalues
	\begin{equation}\label{eq:SigmaSquareEigenvalue_i}
	\hat{\boldsymbol{\sigma}}^2_i\lvert\ldots ,s_i,m_i,\ldots\rangle_{n_{s_i}}
	=
	\left(a^{s_i}_{i,n_{s_i}}\right)^2 s_i(s_i+1)
	\lvert\ldots ,s_i,m_i,\ldots\rangle_{n_{s_i}},
	\end{equation}
For a non-coupled
spin the parameter in \eqref{eq:Sigma_i} is unique and hence 
$\hat{\sigma}^\alpha_i\equiv\hat{\mathrm{s}}^{\alpha}_i$.
When the spins of $i$-th and $j$-th magnetic centers are coupled with total spin 
operator $\hat{\mathbf{s}}_{ij}=\hat{\mathbf{s}}_{i}+\hat{\mathbf{s}}_{j}$,
relation \eqref{eq:Sigma_i} transforms
into a more general expression.
The corresponding $\boldsymbol{\sigma}$-operator is $\hat{\boldsymbol{\sigma}}_{ij}$,
with components
\begin{equation}\label{eq:Sigma_ij}
\hat{\sigma}^\alpha_{ij}\lvert\ldots ,s_{ij},m_{ij},\ldots\rangle_{n_{s_{ij}}}
=
a^{s_{ij}}_{ij,n_{s_{ij}}}\hat{\mathrm{s}}^{\alpha_{\phantom{j}}}_{ij} 
\lvert\ldots ,s_{ij},m_{ij},\ldots\rangle_{n_{s_{ij}}},
\end{equation}
with $a^{s_{ij}}_{ij,n_{s_{ij}}} \in \mathbb{R}$.
The $\boldsymbol{\sigma}$ rising and lowering
operators of the considered spin pair then satisfy
\begin{equation}\label{eq:Sigma_LR_ij}
\hat{\sigma}^{\pm}_{ij}\lvert\ldots ,s_{ij},m_{ij},\ldots\rangle_{n_{s_{ij}}}
=
a^{s_{ij}}_{ij,n_{s_{ij}}} \hat{\mathrm{s}}^{{\pm}_{\phantom{i}}}_{ij} 
\lvert \ldots ,s_{ij},m_{ij},\ldots\rangle_{n_{s_{ij}}}.
\end{equation}
Accordingly, for the eigenvalues of $\hat{\boldsymbol{\sigma}}^2_{ij}$ we have
	\begin{equation}\label{eq:SigmaSquareEigenvalue_ij}
	\left( a^{s_{ij}}_{ij,n_{s_{ij}}}\right)^2 s_{ij}(s_{ij}+1).
	\end{equation}
Unlike \eqref{eq:Sigma_i}, the $\boldsymbol{\sigma}$-operators associated to each individual 
spin, in the spin pair, share the same parameter such that
\begin{equation}\label{eq:Sigma_k}
\hat{\sigma}^{\alpha}_i\lvert\ldots ,s_{ij},m_{ij},\ldots\rangle_{n_{s_{ij}}}
=
a^{s_{ij}}_{ij,n_{s_{ij}}} \hat{\mathrm{s}}^{\alpha_{\phantom{j}}}_i 
\lvert\ldots ,s_{ij},m_{ij},\ldots\rangle_{n_{s_{ij}}}.
\end{equation}

When the total spin quantum number is a good quantum number, relations
\eqref{eq:Sigma_ij}--\eqref{eq:Sigma_k} have to be rewritten
accordingly
\begin{equation*}
a^{s_{ij}}_{ij,n_{s_{ij}}}\to a^{s_{ij},s}_{ij,n_{s_{ij},s}}.
\end{equation*}
For example, relation \eqref{eq:Sigma_k} will be given by
\begin{equation}\label{eq:Sigma_k_s}
\hat{\sigma}^{\alpha}_i\lvert\ldots ,s_{ij},\ldots,s,m\rangle_{n_{{s_{ij},s}}}
=
a^{s_{ij},s}_{ij,n_{{s_{ij},s}}} \hat{\mathrm{s}}^{\alpha_{\phantom{j}}}_i 
\lvert\ldots ,s_{ij},\ldots,s,m\rangle_{n_{{s_{ij},s}}}.
\end{equation}

In general, the coefficients in \eqref{eq:Sigma_k} and \eqref{eq:Sigma_k_s} are
a function of two types of parameters. In particular, we have
\begin{subequations}\label{eq:Constraint_ij}
	\begin{equation}\label{eq:as_ij}
	a^{s_{ij}}_{ij,n_{s_{ij}}}=
	h^{s_{ij}}_{ij}c^{s_{ij}}_{ij,n_{s_{ij}}}
	\end{equation}
	and
	\begin{equation}\label{eq:as_ij_s}
	a^{s_{ij},s}_{ij,n_{s_{ij},s}}=
	h^{s}_{ij}c^{s_{ij},s}_{ij,n_{s_{ij},s}},
	\end{equation}
\end{subequations}
where the ``$c$'' parameters have to account for all transitions predicted in 
\eqref{eq:GenericEnergyTransit} and the ``$h$'' parameters the variations in energy level sequence 
due to the indirect action of externally applied magnetic field, see Sec \ref{sec:KI}.
Therefore, the ``$c$--spectroscopic'' parameters
can be fitted to spectroscopic data for $|\mathbf{B}|=0$ 
and the values of ``$h$--field'' parameters can be fixed 
with respect to the magnetization or magnetic susceptibility measurements
\cite{georgiev_magnetization_2020}.
We would like to point out that the field parameters are not explicit 
functions of the applied magnetic field. They account only for the maximum rate
of broadness in energy gaps \eqref{eq:GenericEnergyTransit} 
related to a certain spin multiplet.  

Generally, ``$c$'' parameters are 
functions of the coulomb, hopping and direct exchange integrals.
In the case of trivial bridging structure and 
uniform electrons' distribution, 
$n_{s_{ij}}=1$, $n_{s_{ij},s}=1$ and the corresponding parameters will be equal to unity.
In other words, the spatial component of the state functions
\eqref{eq:OrbitalSymmetrizedBasisStates} will be unique and so the energy gap 
related to indexes $\tau$ and $n_{\tau,s_{\tau},s}$.
Such case is expected to correspond to sharp peaks in the magnetic spectrum. 
Hence the Hamiltonian \eqref{eq:SpinSigmaHamiltonian} reduces to the Heisenberg model.
On the other hand, for all $n_{s_{ij}}\ne n'_{s_{ij}}$, 
the inequality $\big|c^{s_{ij}}_{ij,n_{s_{ij}}}-c^{s_{ij}}_{ij,n'_{s_{ij}}}\big|>0$ 
would have to be considered as a sign for broadened peaks that
might split.

\subsection{The effective $g$-factor}\label{sec:ZH}
	
Taking into account \eqref{eq:G-Factor} and \eqref{eq:GenericEnergyExplic}
for the $\alpha$ component of the operator
\eqref{eq:SigmaZeeman}, we have
\begin{equation}\label{eq:SlikeOperator}
	\hat{S}^\alpha_i\lvert\ldots,s,m\rangle_{n_{\ldots,s}}=
	\mathrm{g}^{\alpha}_{i,s,m}
	\hat{\mathrm{s}}^\alpha_i\lvert\ldots,s,m\rangle_{n_{\ldots,s}}
\end{equation}
where
\begin{equation}\label{eq:G-FactorEff}
	\mathrm{g}^{\alpha}_{i,s,m}=
	\sum_{[\ldots,s_\tau]}\sum_{\tau,n_{\tau,s_\tau,s}}
	\left|c^{(n_{\tau,s_\tau,s})}_{\tau,s_\tau,s,m}\right|^2
	\rho^{(n_{\tau,s_\tau,s})}_{\tau,s_\tau,s,m}
	\langle g^{\alpha}_i\rangle^{(n_{\tau,s_\tau,s})}_{\tau,s_\tau,s}.
\end{equation}
Here, according to \eqref{eq:G-FactorAverage}, we get
\begin{equation*}
	\langle g^{\alpha}_i\rangle^{(n_{\tau,s_\tau,s})}_{\tau,s_\tau,s}=
	\left\langle g_sn_{\alpha}	-
	\lambda\!\!\sum^{}_{j|_{j\ne i}}
	\alpha_{j}\left(\mathbf{r}_{ij}\cdot\mathbf{n}\right)-
	n_{\alpha}\left(\mathbf{r}_{j}\cdot\mathbf{r}_{ij}\right)
	\right\rangle^{(n_{\tau,s_\tau,s})}_{\tau,s_\tau,s},
\end{equation*}
where $\lambda=g_se^2\mu_0/8m_e\pi^2\langle r_{ij}\rangle^{3}$.


\section{Case study: The Spin-one dimer}\label{sec:Dimer}

The oxygen molecule is one among many fundamental examples of spin dimer
molecules. The direct coupling of both oxygen atoms gives rise to a single
transition \cite{masuda_magnetic_2008} emphasizing the key role of the
molecular orbital and the magnetic exchange theories, predicting one
triplet-singlet transition. Another larger dimeric molecule demonstrating
the closer relation between theory and experiment is the copper acetate 
\cite{furrer_magnetic_2010}.
For such trivial bonds and dimeric structure the theory does not predict
any different behavior. However, as it is suggested in Sec. \ref{sec:PHF}, 
in case of complex bridging structure
between the magnetic centers additional effects may be observed. 
Exploring the possible outcomes form the formalism discussed in Sec. 
\ref{sec:PHF} requires a lot of analytical efforts.

\begin{figure}[t!]
\centering
\includegraphics[width=\columnwidth]{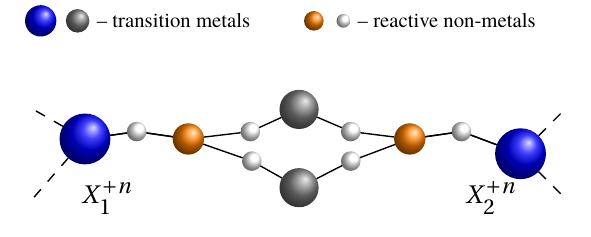}
\caption{Sketch of a spin-one dimeric magnetic molecule with non-trivial exchange bridge. 
The identical spin-one ions $X^{+n}_i$, with $i=1,2$ and $n\in\mathbb{N}$, are colored in blue,
where the letter ``$X$'' refers to magnetic transition metals.
The remaining transition metals (dark-gray balls) belonging to the bridging complex are effectively non-magnetic.
}
\label{fig:Dimer}
\end{figure}

In order to shed more light on the present formalism we consider a fictive spin-one 
dimer molecular magnet with two identical metallic 
centers connected by a non-trivial bridging structure, see Fig. \ref{fig:Dimer}. 
The ground state of the dimer is singlet. 
Another example for the application of the introduced method and spin-sigma Hamiltonian, 
including a fictive spin-half dimer with two non-trivial bridges, is
given in Ref. \cite{georgiev_mexchange_2019}.
Moreover, the present formalism successfully reproduced
the magnetic properties of real compounds,
such as the trimeric spin-half clusters A$_3$Cu$_3$(PO$_4$)$_4$ with A = (Ca, Sr, Pb) \cite{georgiev_trimer_2019}
and the molecular magnet Ni$_4$Mo$_{12}$
\cite{georgiev_epjb_2019,georgiev_magnetization_2020}.

\subsection{State functions and energy levels}\label{sec:DSF}

Let $N$ be the total number of all electrons in the considered dimer 
described by Hamiltonian \eqref{eq:CoulombHamilton}.
Since $N$ is an even number and we consider one exchange bridge,
the total number of orbitals will be given by $n=N/2+2$, where 
we have $n-4$ core and $4$ active molecular orbitals.
Respectively, the index $\tau$ will be unique and given by $\tau=(n-3,n-2),(n-1,n)$. 
We assume the energy of orbitals $(n-1,n)$ to be slightly higher than the energy 
of the pair $(n-3,n-2)$ yet the orbitals within each pair are
energetically equal. 
Therefore, the number of all functions
in \eqref{eq:OrbitalSymmetrizedBasisStates} will depend on the values of 
$v_{\tau,s_{n-3,n-2},s_{n-1,n},s}$, where the 
considered system is characterized by the spin coupling
set $\{s_{n-3,n-2},s_{n-1,n},s,m\}$. 

With respect to the total spin quantum number we distinguish three levels $s=0,1,2$.
The singlet level consists of two spin-multiplets $\{0,0,0,0\}$ and
$\{1,1,0,0\}$.
Accounting for all energetically
distinct spin-orbitals configurations, 
described by the functions \eqref{eq:OrbitalSymmetrizedBasisStates}, the 
multiplet $\{0,0,0,0\}$ appears as $9$-fold degenerate. Thus,
we have $v_{\tau,0,0,0}=1,2\ldots,9$ and $v_{\tau,1,1,0}=1$, respectively.
In the case of constructive 
interference between double occupied $(n-3)$-th and $(n-2)$-th molecular orbitals, 
from all nine multiplets $\{0,0,0,0\}$, we distinguish three states
\begin{subequations}\label{eq:DimerSinglets}
	\begin{align}
	\Psi^{\tau,1}_{0,0,0,0}(\mathbf{r}_1,\ldots,\mathbf{r}_N)=
	&
	\sum_{P_{\mathbf{r}_1\ldots\mathbf{r}_{N}}} \!\!\!
	\frac{c_{\mathbf{r}_{1},\ldots,\mathbf{r}_{N}}}{\sqrt{2^{-\frac{N}{2}}N!}}
	\prod_{i=1}^{\frac{N}{2}-2}
		\Phi^i_{0,0}(\mathbf{r}_{2i-1},\mathbf{r}_{2i}) 
	\nonumber \\
	& \times
	\Phi^{n-3,n-2}_{0,0}(\mathbf{r}_{N-3},\mathbf{r}_{N-2})
	\Phi^{n-1,n}_{0,0}(\mathbf{r}_{N-1},\mathbf{r}_{N}),
	\end{align}
	\begin{align}
	\Psi^{\tau,2}_{0,0,0,0}(\mathbf{r}_1,\ldots,\mathbf{r}_N)=
		&
	\sum_{P_{\mathbf{r}_1\ldots\mathbf{r}_{N}}} \!\!\!
	\frac{c_{\mathbf{r}_{1},\ldots,\mathbf{r}_{N}}}{\sqrt{2^{-\frac{N}{2}}N!}}
	\prod_{i=1}^{\frac{N}{2}-2}
	\Phi^i_{0,0}(\mathbf{r}_{2i-1},\mathbf{r}_{2i}) 
	\nonumber \\
	& \times
	\Phi^{n-3,n-2}_{0,0}(\mathbf{r}_{N-3},\mathbf{r}_{N-2})
	\varPhi^{n-1,n}_{0,0}(\mathbf{r}_{N-1},\mathbf{r}_{N}),
	\end{align}	
	\begin{align}
	\Psi^{\tau,3}_{0,0,0,0}(\mathbf{r}_1,\ldots,\mathbf{r}_N)=
	&
	\sum_{P_{\mathbf{r}_1\ldots\mathbf{r}_{N}}} \!\!\!
	\frac{c_{\mathbf{r}_{1},\ldots,\mathbf{r}_{N}}}{\sqrt{2^{-\frac{N}{2}}N!}}
	\prod_{i=1}^{\frac{N}{2}-2}
	\Phi^i_{0,0}(\mathbf{r}_{2i-1},\mathbf{r}_{2i}) 
	\nonumber \\
	& \times
	\Phi^{n-3,n-2}_{0,0}(\mathbf{r}_{N-3},\mathbf{r}_{N-2})
	\Psi^{n-1,n}_{0,0}(\mathbf{r}_{N-1},\mathbf{r}_{N}),
	\end{align}
\end{subequations}
where the coefficients $c_{\mathbf{r}_{1},\ldots,\mathbf{r}_{N}}$ 
account for the antisymmetry of the functions and the sum runs over the permutation 
of all electrons occupying orbitals of different energy. Here, the permutation
$\Phi^n_{0,0}(\mathbf{r}_i,\mathbf{r}_j)\to\Phi^n_{0,0}(\mathbf{r}_j,\mathbf{r}_i)$
is not included, see \eqref{eq:App4Electrons}.
Further, according to \eqref{eq:MOrbitals} and in terms of up and down notations the 
functions in the summands are given by
\begin{equation*}
	\Phi^n_{0,0}(\mathbf{r}_i,\mathbf{r}_j)=\frac{1}{\sqrt{2}}
	\big(
	\phi_{n,\uparrow}(\mathbf{r}_i)\phi_{n,\downarrow}(\mathbf{r}_j)
	-
	\phi_{n,\downarrow}(\mathbf{r}_i)\phi_{n,\uparrow}(\mathbf{r}_j)
	\big),
\end{equation*}
where $s_n=0$, $m_n=0$,
\begin{align*}
	\Phi^{n',n}_{0,0}(\mathbf{r}_i,\mathbf{r}_j)=
	&\frac{1}{\sqrt{4}}
	\Big(
	\phi_{n',\uparrow}(\mathbf{r}_i)\phi_{n',\downarrow}(\mathbf{r}_j)
	\\ & -
	\phi_{n',\downarrow}(\mathbf{r}_i)\phi_{n',\uparrow}(\mathbf{r}_j)
	\\ & +
	\phi_{n,\uparrow}(\mathbf{r}_i)\phi_{n,\downarrow}(\mathbf{r}_j)
	 -
	\phi_{n,\downarrow}(\mathbf{r}_i)\phi_{n,\uparrow}(\mathbf{r}_j)
	\Big),
\end{align*}
with $s_{n',n}=0$ and $m_{n',n}=0$ the constructive superposition 
between doubly occupied $n'$-th and $n$-th orbitals,
\begin{align*}
	\varPhi^{n',n}_{0,0}(\mathbf{r}_i,\mathbf{r}_j)=
	&\frac{1}{\sqrt{4}}
	\Big(
	\phi_{n',\uparrow}(\mathbf{r}_i)\phi_{n',\downarrow}(\mathbf{r}_j)
	\\ & -
	\phi_{n',\downarrow}(\mathbf{r}_i)\phi_{n',\uparrow}(\mathbf{r}_j)
	\\ & -
	\phi_{n,\uparrow}(\mathbf{r}_i)\phi_{n,\downarrow}(\mathbf{r}_j)
	 +
	\phi_{n,\downarrow}(\mathbf{r}_i)\phi_{n,\uparrow}(\mathbf{r}_j)
	\Big),
\end{align*}
is the destructive counterpart and 
\begin{align*}
	\Psi^{n',n}_{0,0}(\mathbf{r}_i,\mathbf{r}_j)=
	&\frac{1}{\sqrt{4}}
	\Big(
	\phi_{n',\uparrow}(\mathbf{r}_i)\phi_{n,\downarrow}(\mathbf{r}_j)
	-
	\phi_{n',\downarrow}(\mathbf{r}_i)\phi_{n,\uparrow}(\mathbf{r}_j)
	\\ & +
	\phi_{n,\uparrow}(\mathbf{r}_i)\phi_{n',\downarrow}(\mathbf{r}_j)
	 -
	\phi_{n,\downarrow}(\mathbf{r}_i)\phi_{n',\uparrow}(\mathbf{r}_j)
	\Big),
\end{align*}
corresponds to the same local singlet with half-filled $n'$-th and $n$-th orbitals.

The triplet level is build up from three spin-multiplets, 
$\{0,1,1,m\}$, $\{1,0,1,m\}$ and $\{1,1,1,m\}$, where  
$v_{\tau,0,1,1}=1,2,3$, $v_{\tau,1,0,1}=1,2,3$ and $v_{\tau,1,1,1}=1$, respectively.
As an example, consider the triplet $\{0,1,1,+1\}$ with $s_{n-3,n-2}=0$, 
$s_{n-1,n}\equiv s=1$ and $m_{n-1,n}\equiv m=+1$.
Thus, we have 
\begin{align}\label{eq:DimerTriplet}
	\Psi^{\tau,1}_{0,1,1,+1}(\mathbf{r}_1,\ldots,\mathbf{r}_N)=
	&
	\sum_{P_{\mathbf{r}_1\ldots\mathbf{r}_{N}}} \!\!\!
	\frac{c_{\mathbf{r}_{1},\ldots,\mathbf{r}_{N}}}{\sqrt{2^{-\frac{N}{2}}N!}}
	\prod_{i=1}^{\frac{N}{2}-2}
	\Phi^i_{0,0}(\mathbf{r}_{2i-1},\mathbf{r}_{2i}) 
	\nonumber \\
	& \!\!\times\!\!
	\Phi^{n-3,n-2}_{0,0}(\mathbf{r}_{N-3},\mathbf{r}_{N-2})
	\Psi^{n-1,n}_{1,+1}(\mathbf{r}_{N-1},\mathbf{r}_{N}),
\end{align}
\begin{align*}
	\Psi^{\tau,2}_{0,1,1,+1}(\mathbf{r}_1,\ldots,\mathbf{r}_N)=
	&
	\sum_{P_{\mathbf{r}_1\ldots\mathbf{r}_{N}}} \!\!\!
	\frac{c_{\mathbf{r}_{1},\ldots,\mathbf{r}_{N}}}{\sqrt{2^{-\frac{N}{2}}N!}}
	\prod_{i=1}^{\frac{N}{2}-2}
	\Phi^i_{0,0}(\mathbf{r}_{2i-1},\mathbf{r}_{2i}) 
	\nonumber \\
	& \!\!\times\!\!
	\varPhi^{n-3,n-2}_{0,0}(\mathbf{r}_{N-3},\mathbf{r}_{N-2})
	\Psi^{n-1,n}_{1,+1}(\mathbf{r}_{N-1},\mathbf{r}_{N}),
\end{align*}
\begin{align*}
	\Psi^{\tau,3}_{0,1,1,+1}(\mathbf{r}_1,\ldots,\mathbf{r}_N)=
	&
	\sum_{P_{\mathbf{r}_1\ldots\mathbf{r}_{N}}} \!\!\!
	\frac{c_{\mathbf{r}_{1},\ldots,\mathbf{r}_{N}}}{\sqrt{2^{-\frac{N}{2}}N!}}
	\prod_{i=1}^{\frac{N}{2}-2}
	\Phi^i_{0,0}(\mathbf{r}_{2i-1},\mathbf{r}_{2i}) 
	\nonumber \\
	& \!\!\!\times\!\!
	\Psi^{n-3,n-2}_{0,0}(\mathbf{r}_{N-3},\mathbf{r}_{N-2})
	\Psi^{n-1,n}_{1,+1}(\mathbf{r}_{N-1},\mathbf{r}_{N}),
\end{align*}
where
\begin{align*}
	\Psi^{n',n}_{1,+1}(\mathbf{r}_i,\mathbf{r}_j)=
	&\frac{1}{\sqrt{2}}
	\Big(
	\phi_{n',\uparrow}(\mathbf{r}_i)\phi_{n,\uparrow}(\mathbf{r}_j)
	-
	\phi_{n,\uparrow}(\mathbf{r}_i)\phi_{n',\uparrow}(\mathbf{r}_j)
	\Big).
\end{align*}

The quintet level, $\{1,1,2,m\}$, is unique and it 
is non-degenerate regarding the existing spin-orbitals configurations. 
As a result, the index $v_{\tau,1,1,2}=1$ and the state of maximal spin 
magnetic moment $m=2$, reads
\begin{align*}
	\Psi^{\tau,1}_{1,1,2,+2}(\mathbf{r}_1,\ldots,\mathbf{r}_N)=
	&
	\sum_{P_{\mathbf{r}_1\ldots\mathbf{r}_{N}}} \!\!\!
	\frac{c_{\mathbf{r}_{1},\ldots,\mathbf{r}_{N}}}{\sqrt{2^{-\frac{N}{2}}N!}}
	\prod_{i=1}^{\frac{N}{2}-2}
	\Phi^i_{0,0}(\mathbf{r}_{2i-1},\mathbf{r}_{2i}) 
	\nonumber \\
	& \!\!\!\times\!\!
	\Psi^{n-3,n-2}_{1,+1}(\mathbf{r}_{N-3},\mathbf{r}_{N-2})
	\Psi^{n-1,n}_{1,+1}(\mathbf{r}_{N-1},\mathbf{r}_{N}).
\end{align*}

Constructing the effective matrix elements with the aid of \eqref{eq:GeneralEffHam}
and accounting for the integrals given in Sec. \ref{sec:KI},
with respect to the singlet states \eqref{eq:DimerSinglets}, we get
\begin{subequations}\label{eq:DimerSingletElement}
	\begin{equation}
	\big\langle\Psi^{\tau,1}_{0,0,0,0}\big\rvert
	\hat{H}_{\mathrm{eff}}
	\big\lvert\Psi^{\tau,1}_{0,0,0,0}\big\rangle=
	U_\tau+D_\tau,
	\end{equation}
	\begin{equation}
	\big\langle\Psi^{\tau,2}_{0,0,0,0}\big\rvert
	\hat{H}_{\mathrm{eff}}
	\big\lvert\Psi^{\tau,2}_{0,0,0,0}\big\rangle=
	U_\tau-D_\tau,
	\end{equation}
	\begin{equation}
	\big\langle\Psi^{\tau,3}_{0,0,0,0}\big\rvert
	\hat{H}_{\mathrm{eff}}
	\big\lvert\Psi^{\tau,3}_{0,0,0,0}\big\rangle=
	V_\tau+D_\tau,
	\end{equation}
\end{subequations}
respectively. Furthermore, using the triplet state 
\eqref{eq:DimerTriplet}, for $m=0,\pm1$ we have
\begin{align}\label{eq:DimerTripletElements}
	\big\langle\Psi^{\tau,1}_{0,1,1,m}\big\rvert
	\hat{H}_{\mathrm{eff}}
	\big\lvert\Psi^{\tau,1}_{0,1,1,m}\big\rangle
	= &
	V_\tau-D_\tau
	-\mu_{\mathrm{B}}B m \sum^{N}_{i=1}
	\langle g^{\alpha}_i\rangle^{(1)}_{\tau,0,1,1}.
\end{align}
Working with \eqref{eq:DimerSingletElement} and
\eqref{eq:DimerTripletElements}, one has to bear in mind that
\eqref{eq:LowOrbitalAverageEnergy} is rewritten according to the 
considered case and for example $U_\tau$ is a function of the single orbital
integrals $U_{n-3},U_{n-2},U_{n-1}$ and $U_n$.

Computing the eigenvalues in \eqref{eq:BeforeGenericStateFunction} we obtain 
$9$ energy levels associated to the total singlet $s=0$, 
where $n_{\tau,0,0,0}=1,2,\ldots,9$ and $n_{\tau,1,1,0}=1$. Further, we get
$21$ triplet levels corresponding to the indexes $n_{\tau,0,1,1}=1,2,3$, 
$n_{\tau,1,0,1}=1,2,3$ and $n_{\tau,1,1,1}=1$ for all $m=0,\pm1$.
As an example, since the triplet $\{1,1,1,m\}$ is unique, 
from \eqref{eq:StatesAfterDiagonalization} we get
$\lvert\Omega^{\tau,1}_{1,1,1,m}\rangle\equiv\lvert\Psi^{\tau,1}_{1,1,1,m}\rangle$.
Consequently, the corresponding eigenvalues from
\eqref{eq:BeforeGenericStateFunction_a} and \eqref{eq:BeforeGenericStateFunction_b}
read
\begin{equation*}
	E^{\mathrm{f},(1)}_{\tau,1,1,1,m}=V'_\tau-D'_\tau
\end{equation*}
and
\begin{equation*}
	\mathcal{E}^{(1)}_{\tau,1,1,1,m}=
	-\mu_{\mathrm{B}}B m \sum^{N}_{i=1}
	\langle g^{\alpha}_i\rangle^{(1)}_{\tau,1,1,1},
\end{equation*}
respectively, where due to the different spin configuration related to the orbitals $n-3$ 
and $n-2$, we have $V'_\tau\neq V_\tau$ and $D'_\tau\neq D_\tau$. 

The quintet level consists of five energy levels with $n_{\tau,1,1,2}=1$ and $m=0,\pm1,\pm2$.

\subsection{Energy spectrum}\label{sec:DES}

Constructing the effective total spin space we take into account 
\eqref{eq:GenericStateFunction}. Since index 
$\tau$ is unique the sum runs over all values of $n_{\tau,s_{n-3,n-2},s_{n-1,n},s}$
for each spin multiplet $\{s_{n-3,n-2},s_{n-1,n},s,m\}$.
Thus, for all $m=0,\pm1$, the triplet states read 
\begin{align*}
	\lvert 1,m\rangle= &
	c^{(1)}_{1,1,1,m}
	\big\lvert\Omega^{\tau,n_{\tau,1,1,1}}_{1,1,1,m}\big\rangle
	+\!\!\!\!
	\sum^{3}_{n_{\tau,0,1,1}=1}
	c^{(n_{\tau,0,1,1})}_{0,1,1,m}
	\big\lvert\Omega^{\tau,n_{\tau,0,1,1}}_{0,1,1,m}\big\rangle
\nonumber \\
	& +
	\sum^{3}_{n_{\tau,1,0,1}=1}
	c^{(n_{\tau,1,0,1})}_{1,0,1,m}
	\big\lvert\Omega^{\tau,n_{\tau,1,0,1}}_{1,0,1,m}\big\rangle,
\end{align*}
where according to \eqref{eq:GenericEnergyExplic} the corresponding triplet 
energy is given by
\begin{subequations}\label{eq:DEnergyExplic}
	\begin{align}\label{eq:DTripletEnergyExplic}
	E^{\mathrm{f}}_{1,m}=&
	\big|c^{(1)}_{1,1,1,m}\big|^2\rho^{(1)}_{1,1,1,m}\varSigma^{\mathrm{f},(1)}_{1,1,1,m}
\nonumber \\
	& + \sum^{3}_{n_{\tau,0,1,1}=1}\big|c^{(n_{\tau,0,1,1})}_{0,1,1,m}\big|^2
	\rho^{(n_{\tau,0,1,1})}_{0,1,1,m}\varSigma^{\mathrm{f},(n_{\tau,0,1,1})}_{0,1,1,m}
\nonumber \\
	& + \sum^{3}_{n_{\tau,1,0,1}=1}\big|c^{(n_{\tau,1,0,1})}_{1,0,1,m}\big|^2
	\rho^{(n_{\tau,1,0,1})}_{1,0,1,m}\varSigma^{\mathrm{f},(n_{\tau,1,0,1})}_{1,0,1,m}.
\end{align}
For the sake of clarity we write down the singlet energy,
\begin{align}\label{eq:DSingletEnergyExplic}
	E^{\mathrm{f}}_{0,0}=&
	\big|c^{(1)}_{1,1,0,0}\big|^2\rho^{(1)}_{1,1,0,0}\varSigma^{\mathrm{f},(1)}_{1,1,0,0}
\nonumber \\
	& + \sum^{9}_{n_{\tau,0,0,0}=1}\big|c^{(n_{\tau,0,0,0})}_{0,0,0,0}\big|^2
	\rho^{(n_{\tau,0,0,0})}_{0,0,0,0}\varSigma^{\mathrm{f},(n_{\tau,0,0,0})}_{0,0,0,0}.
\end{align}

The quintet level is unique. For all $m$, $c^{(1)}_{1,1,2,m}=1$ and hence
\begin{equation}\label{eq:DQuintetEnergyExplic}
	E^{\mathrm{f}}_{2,m}=\rho^{(1)}_{1,1,2,m}\varSigma^{\mathrm{f},(1)}_{1,1,2,m}.
\end{equation}
\end{subequations}
In the case $|\mathbf{B}|=0$, we have $E^{\mathrm{f}}_{s,m}\to E_{s,m}$.

\subsection{Magnetic transitions}\label{sec:DMT}
  
In the absence of an external magnetic field and at $T\to0$ 
the dimer is in its ground state with energy $E_{0,0}$. The first excited 
level is the triplet with energy $E_{1,m}$, where $m=0,\pm1$.   

Assume measuring the dimer's magnetic spectrum, we perform an
inelastic neutron scattering experiment. 
Applying the corresponding selection rules under the considered spin coupling scheme,
we take into account that for the allowed energy transitions the local spin quantum 
numbers $s_{n-3,n-2}$ and $s_{n-1,n}$ cannot be simultaneously changed. 
Accordingly, with respect to the quantum probabilities in \eqref{eq:DTripletEnergyExplic} 
and \eqref{eq:DSingletEnergyExplic}, the applied variational method predicts
one broadened low-temperature peak in the magnetic spectrum, centered at 
$E_{1,\Delta m}=E_{1,m}-E_{0,0}$. 
In particular, the obtained peak will be a product of three slightly
energetically different
magnetic excitations. One associated to the energy transition with $\Delta s_{n-3,n-2}=0$, $\Delta s_{n-1,n}=0$
and $\Delta s=1$, a second with $\Delta s_{n-3,n-2}=0$, $\Delta
s_{n-1,n}=1$, $\Delta s=1$ and a third 
corresponding to the energy transition for which $\Delta s_{n-3,n-2}=1$, $\Delta s_{n-1,n}=0$, $\Delta s=1$.
Thus, depending on the difference in energies of both orbital 
pairs $(n-3,n-2)$ and $(n-1,n)$ we may expect splitting of the peak.

Theoretically, a high-temperature magnetic transition is also allowed. 
The center of the corresponding peak will be characterized by the energy 
$E_{1,\Delta m}=E_{2,m'}-E_{1,m}$, for all $m'=0,\pm1,\pm2$.
However, due to the higher temperature, features like broadening and splitting may 
not be pronounced. 

Henceforth, we assume that for $T\to0$ the probability for observing 
the transition with $\Delta s_{n-3,n-2}=0$, $\Delta s_{n-1,n}=0$ and 
$\Delta s=1$ is negligible and hence the low temperature peak is build of two
different by intensity peaks, the first centered at
\begin{subequations}\label{eq:DTransitionEnergies}
	\begin{align}\label{eq:DTransitionEnergies_a}
	E^{(1)}_{1,\Delta m}=&
	\sum^{3}_{n_{\tau,0,1,1}=1}\big|c^{(n_{\tau,0,1,1})}_{0,1,1,m}\big|^2
	\rho^{(n_{\tau,0,1,1})}_{0,1,1,m}\varSigma^{(n_{\tau,0,1,1})}_{0,1,1,m}
\nonumber \\ 
	& -
	\sum^{9}_{n_{\tau,0,0,0}=1}\big|c^{(n_{\tau,0,0,0})}_{0,0,0,0}\big|^2
	\rho^{(n_{\tau,0,0,0})}_{0,0,0,0}\varSigma^{(n_{\tau,0,0,0})}_{0,0,0,0}
\end{align}
and the second at
\begin{align}\label{eq:DTransitionEnergies_b}
	E^{(2)}_{1,\Delta m}=&
	\sum^{3}_{n_{\tau,1,0,1}=1}\big|c^{(n_{\tau,1,0,1})}_{1,0,1,m}\big|^2
	\rho^{(n_{\tau,1,0,1})}_{1,0,1,m}\varSigma^{(n_{\tau,1,0,1})}_{1,0,1,m}
	\nonumber \\ 
	& -
	\sum^{9}_{n_{\tau,0,0,0}=1}\big|c^{(n_{\tau,0,0,0})}_{0,0,0,0}\big|^2
	\rho^{(n_{\tau,0,0,0})}_{0,0,0,0}\varSigma^{(n_{\tau,0,0,0})}_{0,0,0,0},
\end{align}
where $E^{(2)}_{1,\Delta m}>E^{(1)}_{1,\Delta m}$ and $\Delta m=m$.
Furthermore, let for all $n_{\tau,0,1,1},n_{\tau,1,0,1}=1,2,3$ and $T>0$, we have 
$$
\rho^{(1)}_{1,1,1,m}\gg\rho^{(n_{\tau,0,1,1})}_{0,1,1,m},\rho^{(n_{\tau,1,0,1})}_{1,0,1,m},
$$
whereat the probability of observing a high-temperature magnetic excitation
is related only to the transition with energy 
\begin{equation}
	E^{(3)}_{1,\Delta m}=\rho^{(1)}_{1,1,2,m'}\varSigma^{(1)}_{1,1,2,m'}
	-\big|c^{(1)}_{1,1,1,m}\big|^2\rho^{(1)}_{1,1,1,m}\varSigma^{(1)}_{1,1,1,m},
\end{equation}
for all $\Delta m$.
\end{subequations}
In addition, we imply that for each magnetic center the total effective $g$-factor 
in \eqref{eq:G-FactorEff} is taken within the low-field limit, such that 
$c^{(n_{\tau,s_\tau,s})}_{\tau,s_\tau,s,m}\simeq c^{(n_{\tau,s_\tau,s})}_{\tau,s_\tau,s,m'}$
and $\rho^{(n_{\tau,s_\tau,s})}_{\tau,s_\tau,s,m}\simeq \rho^{(n_{\tau,s_\tau,s})}_{\tau,s_\tau,s,m'}$,
where $m\ne m'$.

\begin{figure}[t!]
	\centering
	\includegraphics[scale=1]{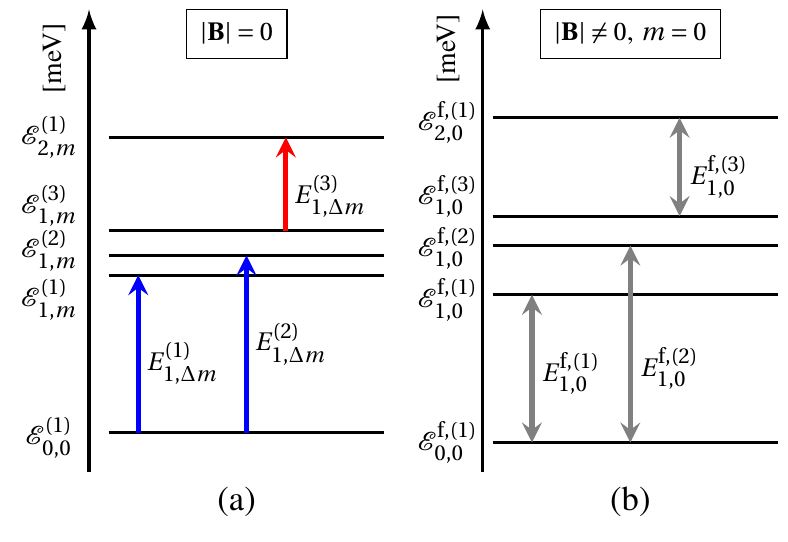}
	\caption{Energy spectrum of the considered spin-one dimer: 
		(a) in the absence of magnetic field; 
		(b) the energy levels of the non-magnetic states 
		in the presence of magnetic field. 
		The blue and red arrows depict the low-temperature and 
		high-temperature magnetic transitions, respectively, see
		\eqref{eq:DLowTransitions} and \eqref{eq:DHighTransitions}. 
		The gray arrows show the energy level's shifting due 
		to the indirect action of $\mathbf{B}$, indicated by the 
		``$h$-field'' parameters in \eqref{eq:Dh}. 
		}
	\label{fig:dspectrum}
\end{figure}

\subsection{The dimer spin-sigma Hamiltonian}\label{sec:DSSH}

In order to construct an adequate energy spectrum that capture the physics behind the obtained 
variational energy spectrum discussed in Sec. \ref{sec:DES}, account for the transitions 
\eqref{eq:DTransitionEnergies} and simplify the computation of
any relevant magnetic
observables, we consider the spin-sigma Hamiltonian
\eqref{eq:TSpinSigmaHamiltonian}, that reads
\begin{equation*}
	\hat{\mathcal{H}}= J 
	\big(
	\hat{\boldsymbol{\sigma}}_1\cdot\hat{\mathbf{s}}_2
	+
	\hat{\boldsymbol{\sigma}}_2\cdot\hat{\mathbf{s}}_1
	\big)
	-
	\mu_{\mathrm{B}}B\Big(\hat{S}^z_1+\hat{S}^z_2\Big),
\end{equation*}
where $J_{12}=J_{21}=J$ and we select the $z$ axis as a quantization axis. 
Both $\boldsymbol{\sigma}$-operators share the same coefficients 
$a_{s,n_s}=h_sc_{s,n_s}$ and hence obey \eqref{eq:Sigma_k}, such that 
$|s_2+s_1|\geq s\geq|s_2-s_1|$ is the pair's total spin quantum number
and $s_i$ is the spin quantum number of the $i$-th effective
magnetic center. 
Further, since both magnetic ions are indistinguishable and the $g$-factor is 
calculated within a low-temperature and low-field regime,
we have $\mathrm{g}^z_{1,s,m}=\mathrm{g}^z_{2,s,m}\to\mathrm{g}^z_{s}$. Therefore, 
using \eqref{eq:Constraint_ij} and \eqref{eq:SlikeOperator}, we get
\begin{equation*}
	\hat{\mathcal{H}}\lvert s,m\rangle_{n_s}=
	\mathcal{E}^{\mathrm{f},(n_{s})}_{s,m}\lvert s,m\rangle_{n_s},
\end{equation*}
where 
\begin{align}\label{eq:DEnSpectrum}
	\mathcal{E}^{\mathrm{f},(n_{s})}_{s,m}=
	& \frac{Jh_sc_{s,n_s}}{2}
	\big(
	s(s+1)-s_1(s_1+1)-s_2(s_2+1)
	\big)
\nonumber \\
	&
	-m\mu_{\mathrm{B}}\mathrm{g}^z_{s}B.
\end{align}

Since all transitions given in \eqref{eq:DTransitionEnergies} 
are related to the triplet state and the quintet state is unique, it follows 
that $n_0=1$, $n_2=1$ and $n_1=1,2,3$, with $c_{0,1}=1$ and $c_{2,1}=1$.
Hence, we are left with three field parameters 
$\{h_s\}^3_{s=1}$ and three spectroscopic parameters $\{c_{1,n_1}\}^3_{n_1=1}$.
The spectrum \eqref{eq:DEnSpectrum} consists of fifteen energy levels
\begin{equation}
	\begin{array}{ll}
		\mathcal{E}^{\mathrm{f},(1)}_{2,m}=Jh_2-m\mu_{\mathrm{B}}\mathrm{g}^z_{2}B, 
		& \mathcal{E}^{\mathrm{f},(3)}_{1,m}=-Jh_1c_{1,3}-m\mu_{\mathrm{B}}\mathrm{g}^z_{1}B,
		\\ [0.25cm]
		\mathcal{E}^{\mathrm{f},(1)}_{1,m}=-Jh_1c_{1,1}-m\mu_{\mathrm{B}}\mathrm{g}^z_{1}B, 
		& \mathcal{E}^{\mathrm{f},(1)}_{0,0}=-2Jh_0,
		\\ [0.25cm]
		\mathcal{E}^{\mathrm{f},(2)}_{1,m}=-Jh_1c_{1,2}-m\mu_{\mathrm{B}}\mathrm{g}^z_{1}B.
		&  {}
	\end{array}
\end{equation}
In the absence of external magnetic field $h_s=1$, for all $s$ and therefore
$\mathcal{E}^{\mathrm{f},(1)}_{1,m}\to\mathcal{E}^{(1)}_{1,m}$, where 
$\mathcal{E}^{(1)}_{1,m}=-Jc_{1,1}$.

According to \eqref{eq:DTransitionEnergies}, we observe two low-temperature magnetic 
excitations with energies
\begin{equation}\label{eq:DLowTransitions}
	\mathcal{E}^{(1)}_{1,m}-\mathcal{E}^{(1)}_{0,0}
	=E^{(1)}_{1,\Delta m},
\qquad
	\mathcal{E}^{(2)}_{1,m}-\mathcal{E}^{(1)}_{0,0}
	=E^{(2)}_{1,\Delta m}
\end{equation}
and a high-temperature one, with energy
\begin{equation}\label{eq:DHighTransitions}
	\mathcal{E}^{(1)}_{2,m^{}_{}}-\mathcal{E}^{(3)}_{1,m'}
	=E^{(3)}_{1,\Delta m}.
\end{equation}
Determining the spectroscopic parameters, from \eqref{eq:DLowTransitions} 
and \eqref{eq:DHighTransitions}, we obtain
\begin{equation}
	J=E^{(1)}_{1,\Delta m},
\quad
	c_{1,1}=1,
\quad
	c_{1,2}=2-\frac{E^{(2)}_{1,\Delta m}}{E^{(1)}_{1,\Delta m}}
\end{equation}
and
\begin{equation}
	c_{1,3}=\frac{E^{(3)}_{1,\Delta m}}{E^{(1)}_{1,\Delta m}}-1,
\end{equation}
respectively. The energy spectrum with all transitions are depicted 
on Fig. \ref{fig:dspectrum} (a). 

Obtaining the expressions for the ``$h$--field'' parameters, we take 
into account the energy levels corresponding only to the non-magnetic states, 
i.e. for $m=0$. Thus, in addition to \eqref{eq:DLowTransitions} and 
\eqref{eq:DHighTransitions}, we have
\begin{equation*}
	\mathcal{E}^{\mathrm{f},(1)}_{1,0}-\mathcal{E}^{\mathrm{f},(1)}_{0,0}
	=E^{\mathrm{f},(1)}_{1,0},
\qquad
	\mathcal{E}^{\mathrm{f},(2)}_{1,0}-\mathcal{E}^{\mathrm{f},(1)}_{0,0}
	=E^{(2)}_{1,0}
\end{equation*}
and 
\begin{equation*}
	\mathcal{E}^{\mathrm{f},(1)}_{2,0}-\mathcal{E}^{\mathrm{f},(3)}_{1,0}
	=E^{\mathrm{f},(3)}_{1,0},
\end{equation*}
respectively. As a result, we get
\begin{subequations}\label{eq:Dh}
\begin{equation}
	h_0=\frac{E^{\mathrm{f},(1)}_{1,0}+E^{\mathrm{f},(2)}_{1,0}}
		{4E^{(1)}_{1,\Delta m}}
		+\frac{1}{4}
		\left( 
		3-\frac{E^{(2)}_{1,\Delta m}}{E^{(1)}_{1,\Delta m}}
		\right)
		\frac{E^{\mathrm{f},(1)}_{1,0}-E^{\mathrm{f},(2)}_{1,0}}
		{E^{(1)}_{1,\Delta m}-E^{(2)}_{1,\Delta m}},
\end{equation}
\begin{equation}
	h_1=\frac{E^{\mathrm{f},(1)}_{1,0}-E^{\mathrm{f},(2)}_{1,0}}
		{E^{(1)}_{1,\Delta m}-E^{(2)}_{1,\Delta m}}
\end{equation}
and
\begin{equation}
	h_2=\frac{E^{\mathrm{f},(3)}_{1,0}}{E^{(1)}_{1,\Delta m}}-
	\left( 
	\frac{E^{(3)}_{1,\Delta m}}{E^{(1)}_{1,\Delta m}}-1
	\right)
	\frac{E^{\mathrm{f},(1)}_{1,0}-E^{\mathrm{f},(2)}_{1,0}}
	{E^{(1)}_{1,\Delta m}-E^{(2)}_{1,\Delta m}}.
\end{equation}
\end{subequations}
An illustration of how the external magnetic field may shift the non-magnetic energy 
levels due to the contribution of all diamagnetic and paramagnetic terms
\eqref{eq:PartialHamiltonians} is shown on Fig. \ref{fig:dspectrum} (b). 
This effect is considered as the main reason for the observed broadening 
of the magnetization steps of the molecular magnet Ni$_4$Mo$_{12}$,
discussed in \cite{georgiev_magnetization_2020}


\section{Conclusion}\label{sec:Conclusion}

We present a detailed discussion on the theoretical framework recently
used to study the magnetic properties of the Cu based spin-trimeric
compounds \cite{georgiev_trimer_2019} and Ni$_4$Mo$_{12}$ molecular
magnet \cite{georgiev_epjb_2019,georgiev_magnetization_2020}. The
introduced method, see Sec. \ref{sec:PHF}, is based on the molecular
orbital theory and follows closely the
multi-configurational self-consistent field approach. 
The method produces a generalized energy spectrum that 
accounts for all distinct in energy spin-orbital configurations and
hence nonhomogeneous electrons' distributions that may result form the spatial
structure of the studied magnetic unit. Moreover,
individual orbitals have but minor contributions due to the
delocalization of electrons,
such that the associated anisotropy effects vanish.
Thus, the obtained energy sequence predicts the existence of multiple
possible magnetic excitations that are not related to energy
transitions from the fine or hyper-fine structure in a zero-field
approximation, but rather result from the excitation of electron pairs of different distributions.
Broadened steps and a high-field saturation in magnetization is yet another
possible feature.

Based on all predictions made
within the framework of our method, in Sec. \ref{sec:SSH} 
we propose a spin-like Hamiltonian with an appropriately selected parametrization scheme 
that captures the relevant magnetic features. 
In particular, the spin Hamiltonian includes two class of parameters,
spectroscopic and field parameters, see \eqref{eq:as_ij}.
The ``$c$--spectroscopic'' parameters account for all 
magnetic excitations contained in the
variational energy spectrum and they can be
fitted according to the 
relevant experimentally observed magnetic spectra.
On the other hand, the ``$h$--field'' parameters address the shifting of all energy levels
due to the indirect action of an externally applied magnetic field, see Sec. \ref{sec:KI}.
Their value can be fixed from the magnetization and magnetic susceptibility measurements.

As an example, for the application of the introduced method 
and spin-sigma model, we consider a fictive spin-one 
dimer molecular magnet with non-trivial bridging structure, see Sec. \ref{sec:Dimer}.
We derive the explicit expressions of the variational state functions 
\eqref{eq:OrbitalSymmetrizedBasisStates},
construct the energy spectrum \eqref{eq:GenericEnergyExplic} and discuss all
allowed magnetic excitations associated to the transitions \eqref{eq:GenericEnergyTransit}.
Further, we introduce the dimer's Hamiltonian \eqref{eq:TSpinSigmaHamiltonian} and 
the corresponding eigenvalues.
In other words, we compute the ``$c$--spectroscopic'' and
``$h$--field'' parameters
and hence show how the spin-sigma energy spectrum allows the
derivation of the main features of that relevant to the variational
approach.

Although the spin-like Hamiltonian \eqref{eq:TSpinSigmaHamiltonian} includes only 
a bilinear exchange interaction term
\eqref{eq:SpinSigmaHamiltonian}, it may describe reasonably well the magnetism in
larger and more complex molecular magnets with $n$d or $n$f metal centers.
However, the application of the underlying method proposed is Sec. \ref{sec:PHF}, 
remains restricted to specific variety of 
compounds in which the electrons are not localized around the magnetic ions
and do not belong to a conduction band. 
Therefore, additional terms that account for any perturbation
parameter addressing spin-orbital anisotropy would be inappropriate.

	
\begin{acknowledgments}
This work was supported by the Bulgarian National Science Fund under
grant No KP-06-N38/6 and the National Program ``Young scientists and
postdoctoral researchers'' approved by DCM 577 on 17.08.2018.
\end{acknowledgments}


\appendix*

\section{}

\subsection{Generalized momentum}\label{sec:AppGenMom}

Since for all $\alpha$ and $i$ the operators $\hat{A}_\alpha(\mathbf{r}_i)$ 
and $\hat{p}^\alpha_i$ given in Section \ref{sec:ElMomenta} commute, the $i$-th 
electron's generalized momentum operator reads
\begin{equation*}\label{eq:AppGeneralMomentum}
	\hat{\mathbf{P}}^2_{i}=\hat{\mathbf{p}}^2_{i}
	-2e\hat{\mathbf{A}}_{ext}(\mathbf{r}_i)\cdot\hat{\mathbf{p}}_{i}+
	2e \!\!\!\! \sum_{j|_{j\ne i}}
	\hat{\mathbf{A}}(\mathbf{r}_{ij})\cdot\hat{\mathbf{p}}_{i} 
	+e^2\hat{\mathbf{A}}^2(\mathbf{r}_i),
\end{equation*}
where 
\begin{equation*}
	\hat{\mathbf{A}}^2(\mathbf{r}_i)= \hat{\mathbf{A}}^2_{ext}(\mathbf{r}_i)+
	2\sum_{j|_{j\ne i}}\hat{\mathbf{A}}_{ext}(\mathbf{r}_i)\cdot\hat{\mathbf{A}}(\mathbf{r}_{ij})+
	\left(\sum_{j|_{j\ne i}}\hat{\mathbf{A}}(\mathbf{r}_{ij})\right)^2.
\end{equation*}
In terms of \eqref{eq:ExplicitVectorPotential} we have the diamagnetic term
\begin{subequations}\label{eq:AppExplicidTerms}
	\begin{equation}\label{eq:DiamagTerm}
	\hat{\mathbf{A}}^2_{ext}(\mathbf{r}_i)=\frac{\mathbf{B}^2\mathbf{r}^2_{i}}{4},
	\end{equation}
the paramagnetic terms
	\begin{equation}\label{eq:ParamegTerm}
	\hat{\mathbf{A}}_{ext}(\mathbf{r}_i)\cdot\hat{\mathbf{p}}_{i}
	=
	\frac{\hbar}{2}\hat{\boldsymbol{l}}_i\cdot\mathbf{B},
	\end{equation}
	\begin{equation}\label{eq:AppExplicidTerms_c}
	\hat{\mathbf{A}}(\mathbf{r}_{ij})\cdot\hat{\mathbf{p}}_{i}
	=
	\frac{\mu_0\mu_{\mathrm{B}}\hbar}{4\pi^2\langle r_{ij}\rangle^{3}}
	\left(
	\hat{\boldsymbol{l}}_i\cdot\hat{\boldsymbol{l}}_j+
	g_s
	\hat{\boldsymbol{l}}_i\cdot\hat{\boldsymbol{s}}_j
	\right),
	\end{equation}
	\begin{align}\label{eq:AppExplicidTerms_d}
	\hat{\mathbf{A}}_{ext}(\mathbf{r}_i)\cdot\hat{\mathbf{A}}(\mathbf{r}_{ij})
	 = &
	\frac{\mu_0\mu_{\mathrm{B}}}{8\pi^2\langle r_{ij}\rangle^{3}}
	\bigg[ 
	\left(\mathbf{r}_{ij}\cdot\mathbf{B}\right) 
	\left(\mathbf{r}_{i}\cdot\hat{\boldsymbol{l}}_j\right)
	\nonumber \\
	& -
	\left(\mathbf{r}_{i}\cdot\mathbf{r}_{ij}\right)
	\left(\hat{\boldsymbol{l}}_j\cdot\mathbf{B}\right)
	\bigg]
	\nonumber \\
	& + 
	\frac{\mu_0\mu_{\mathrm{B}}g_s}{8\pi^2\langle r_{ij}\rangle^{3}}
	\bigg[ 
	\left(\mathbf{r}_{ij}\cdot\mathbf{B}\right) 
	\left(\mathbf{r}_{i}\cdot\hat{\boldsymbol{s}}_j\right)
	\nonumber \\
	& -
	\left(\mathbf{r}_{i}\cdot\mathbf{r}_{ij}\right)
	\left(\hat{\boldsymbol{s}}_j\cdot\mathbf{B}\right)
	\bigg]
	\end{align}
\end{subequations}
and 
\begin{align}\label{eq:AppExplicidTerm}
	\left(\sum_{j|_{j\ne i}}\hat{\mathbf{A}}(\mathbf{r}_{ij})\right)^2
	= &
	\frac{\mu^2_0\mu^2_{\mathrm{B}}}{16\pi^4}
	\sum_{j|_{j\ne i}}\Bigg(
	\frac{\mathbf{r}^2_{ij}\hat{\boldsymbol{l}}^2_j}{\langle r_{ij}\rangle^{6}}
	+\sum_{k|_{k\ne j}}\!\!\left(\langle r_{ij}\rangle\langle r_{ik}\rangle\right)^{-3}	
	\nonumber \\
	& \times
	\big[ 
	(\mathbf{r}_{ij}\cdot\mathbf{r}_{ik})
	(\hat{\boldsymbol{l}}_j\cdot\hat{\boldsymbol{l}}_k)
	-
	(\mathbf{r}_{ij}\cdot\hat{\boldsymbol{l}}_k)
	(\mathbf{r}_{ik}\cdot\hat{\boldsymbol{l}}_j)
	\big] 
	\Bigg)
	\nonumber \\
	& +
	\frac{\mu^2_0\mu^2_{\mathrm{B}}g^2_s}{16\pi^4}
	\sum_{j|_{j\ne i}}\Bigg(
	\frac{\mathbf{r}^2_{ij}\hat{\boldsymbol{s}}^2_j}{\langle r_{ij}\rangle^{6}}
	+\!\!\!\!
	\sum_{k|_{k\ne j}}\!\!\!\left(\langle r_{ij}\rangle\langle r_{ik}\rangle\right)^{-3}
	\nonumber \\	
	& \times	
	\big[ 
	(\mathbf{r}_{ij}\cdot\mathbf{r}_{ik})
	(\hat{\boldsymbol{s}}_j\cdot\hat{\boldsymbol{s}}_k)
	-
	(\mathbf{r}_{ij}\cdot\hat{\boldsymbol{s}}_k)
	(\mathbf{r}_{ik}\cdot\hat{\boldsymbol{s}}_j)
	\big] 
	\Bigg) 
	\nonumber \\ 
	& 
	+	\frac{\mu^2_0\mu^2_{\mathrm{B}}g_s}{16\pi^4}
	\sum_{j,k|_{j\ne k\ne i}} \!\!\!\!\!
	\frac{\big\{
		(\mathbf{r}_{ij}\times\hat{\boldsymbol{l}}_j)_\alpha,
		(\mathbf{r}_{ik}\times\hat{\boldsymbol{s}}_k)_\alpha
		\big\}}{\left(\langle r_{ij}\rangle\langle r_{ik}\rangle\right)^{3}},
\end{align}
where we have the unit vector $\mathbf{n}_{ij}$ defined by 
$\mathbf{n}_{ij}=\mathbf{r}_{ij}/\langle r_{ij}\rangle$.
For $|\mathbf{B}|>0$, in comparison to the terms in \eqref{eq:AppExplicidTerms}, 
the average of the operator in \eqref{eq:AppExplicidTerm} can be neglected.
Thus, taking into account that $\mu_{\mathrm{B}}=e\hbar/2m_e$ within the combination of
\eqref{eq:AppExplicidTerms_c} and \eqref{eq:AppExplicidTerms_d}
we find it more convenient to define the operators \eqref{eq:K_a} and
\eqref{eq:K_b},
entering in the expression \eqref{eq:GeneralMomenum} for the $i$-th electron's 
generalized momentum operator.

\subsection{Four electrons state}\label{sec:AppFourElectrons}

In the case of four electrons, $N=4$, with one core $\phi_1$
and two active $\phi_2$, $\phi_3$ molecular orbitals, the index $\tau=(2,3)$
is unique, { the spin quantum number related to the core orbital
$s_1=0$, $s_{(2,3)}\equiv s$ is the total spin quantum number 
since the spin pair is unique and hence $v_{\tau,s_\tau,s}\equiv v_{(2,3),s}$.
We have three linear independent combinations with $v_{(2,3),0}=1,2,3$ associated to the singlet level $s=0$
and one corresponding to the triplet level $v_{(2,3),1}=1$.}
Let us focus on the non-magnetic triplet indicated by { $v_{(2,3),1}=1$}, $s=1$, $m=0$, 
with unique coefficient { $c^1_{2,3}=1$}
and given according to \eqref{eq:OrbitalSymmetrizedBasisStates} as follows
\begin{align}\label{eq:App4Electrons}
	{ \Psi^{1,(2,3)}_{1,0}}(\mathbf{r}_1,\ldots,\mathbf{r}_{4})
	\equiv
	&
	\frac{1}{\sqrt{6}}
	\Big( 
	\Phi^1_{0,0}(\mathbf{r}_1,\mathbf{r}_2)
	\Psi^{2,3}_{1,0}(\mathbf{r}_3,\mathbf{r}_4)
\nonumber \\ 	
	& -
	\Phi^1_{0,0}(\mathbf{r}_1,\mathbf{r}_{3})
	\Psi^{2,3}_{1,0}(\mathbf{r}_2,\mathbf{r}_4)
\nonumber \\ 
	&
	+
	\Phi^1_{0,0}(\mathbf{r}_1,\mathbf{r}_4)
	\Psi^{2,3}_{1,0}(\mathbf{r}_2,\mathbf{r}_3)
\nonumber \\ 
	&
	+
	\Phi^1_{0,0}(\mathbf{r}_2,\mathbf{r}_3)
	\Psi^{2,3}_{1,0}(\mathbf{r}_1,\mathbf{r}_4)
	\nonumber \\
	&
	-
	\Phi^1_{0,0}(\mathbf{r}_2,\mathbf{r}_{4})
	\Psi^{2,3}_{1,0}(\mathbf{r}_1,\mathbf{r}_{3})
\nonumber \\ 
	&
	+
	\Phi^1_{0,0}(\mathbf{r}_3,\mathbf{r}_4)
	\Psi^{2,3}_{1,0}(\mathbf{r}_1,\mathbf{r}_2)		
	\Big),
\end{align}
where with respect to the denotation \eqref{eq:MOrbitals}, for 
$m_i\in\{\uparrow,\downarrow\}$, we have
\[
	\Phi^1_{0,0}(\mathbf{r}_i,\mathbf{r}_j)=\frac{1}{\sqrt{2}}
	\big(
	\phi_{1,\uparrow}(\mathbf{r}_i)\phi_{1,\downarrow}(\mathbf{r}_j)
	-
	\phi_{1,\downarrow}(\mathbf{r}_i)\phi_{1,\uparrow}(\mathbf{r}_j)
	\big)
\]
and
\begin{align*}
	\Psi^{(2,3)}_{1,0}(\mathbf{r}_i,\mathbf{r}_j)
	=&\frac{1}{\sqrt{4}}
	\Big(
	\phi_{2,\uparrow}(\mathbf{r}_i)\phi_{3,\downarrow}(\mathbf{r}_j)
	+
	\phi_{2,\downarrow}(\mathbf{r}_i)\phi_{3,\uparrow}(\mathbf{r}_j)
	\\ & - 
	\phi_{3,\uparrow}(\mathbf{r}_i)\phi_{2,\downarrow}(\mathbf{r}_j)
	-
	\phi_{3,\downarrow}(\mathbf{r}_i)\phi_{2,\uparrow}(\mathbf{r}_j)
	\Big).	
\end{align*}
On the other hand, considering the magnetic triplet $m=+1$, we have
\[
	\Psi^{(2,3)}_{1,+1}(\mathbf{r}_i,\mathbf{r}_j)=\frac{1}{\sqrt{2}}
	\Big(
	\phi_{2,\uparrow}(\mathbf{r}_i)\phi_{3,\uparrow}(\mathbf{r}_j)-
	\phi_{3,\uparrow}(\mathbf{r}_i)\phi_{2,\uparrow}(\mathbf{r}_j)
	\Big).
\]



%

\end{document}